\documentclass[nofootinbib,aps, pra, superscriptaddress, amsmath, amssymb, twocolumn, nofootinbib,floatfix]{revtex4-1}

\usepackage{amsmath, amssymb}
\usepackage{braket}					
\usepackage{graphicx} 				
\usepackage{hyperref}				
\usepackage[usenames, dvipsnames]{color}	
\usepackage[caption=false]{subfig}
\usepackage{mathtools}

\DeclareMathOperator{\tr}{tr} 								
\newcommand{\dif}{ \operatorname{d}\!}				

\DeclareMathOperator\diag{diag}							

\newcommand{\nn}{\nonumber}

\usepackage{soul}

\begin{document}

\preprint{}

\title{Quantum reference frames associated with non-compact groups: the case of translations and boosts, and the role of mass}



\author{Alexander R. H. Smith}
\email[]{a14smith@uwaterloo.ca}
\affiliation{Department of Physics \& Astronomy, University of Waterloo, Waterloo, Ontario N2L 3G1 Canada}
\affiliation{Department of Physics \& Astronomy, Macquarie University, New South Wales 2109, Australia}

\author{Marco Piani}
\affiliation{SUPA and Department of Physics, University of Strathclyde, Glasgow G4 0NG, UK}
\affiliation{Institute for Quantum Computing and Department of Physics \& Astronomy, University of Waterloo, Waterloo, Ontario N2L 3G1, Canada}

\author{Robert B. Mann}
\affiliation{Department of Physics \& Astronomy, University of Waterloo, Waterloo, Ontario N2L 3G1 Canada}
\affiliation{Perimeter Institute for Theoretical Physics, 31 Caroline St. N., Waterloo, Ontario N2L 2Y5, Canada}


\date{\today}

\begin{abstract}
Quantum communication without a shared reference frame or the construction of a relational quantum theory requires the notion of a quantum reference frame. We analyze aspects of quantum reference frames associated with non-compact groups, specifically the group of spatial translations and Galilean boosts. We begin by demonstrating how the usually employed group average, used to dispense of the notion of an external reference frame, leads to unphysical states when applied to reference frames associated with non-compact groups. However, we show that this average does lead naturally to a reduced state on the relative degrees of freedom of a system, which was previously considered by \citet{Angelo:2011}. We then study in detail the informational properties of this reduced state for systems of two and three particles in Gaussian states. 
\end{abstract}


\maketitle

\section{Introduction}
\label{Introduction}

The central lesson of relativity is that all observable quantities are relational: length, time, and energy, which were once thought to be absolute, only have meaning with respect to an observer. The same is true of a quantum state. For example, when we write the quantum state $\ket{\uparrow}$, say up in $z$, what we mean is somebody in a laboratory with an appropriately aligned measuring apparatus will measure a specific outcome. This is the description of a quantum state with respect to a classical object, in this example the macroscopic laboratory. 

This state of affairs is not fully satisfactory, since a quantum system is being described with respect to a classical system, that is, by mixing elements of conceptually different frameworks. If we believe that our world is completely described by quantum mechanics, we should seek a theory in which quantum systems are described with respect to quantum systems. Much work has been done on this subject, known as quantum reference frames \cite{Bartlett:2007}, and it has found applications in quantum interferometry \cite{Jarzyna:2012}, quantum communication \cite{Bartlett:2009}, and cryptography \cite{Kitaev:2004}, as well as offering an explanation of previously postulated superselction rules \cite{Aharonov:1967, Dowling:2006}.

Additionally, treating reference frames quantum mechanically is a crucial step towards the goal of constructing  a relational quantum theory \cite{Rovelli:1991,Rovelli:1996}. By relational it is meant a theory that does not make use of an external reference frame to specify its elements. The main motivation for this is general relativity, which does not use an external reference frame in its construction. It is believed that a theory of quantum gravity will inherent this property, and thus, a theory of quantum gravity will necessarily include a theory of quantum reference frames \cite{Poulin:2006, Pienaar:2016}.

The natural language of reference frames is that of group theory, owing to the fact that the transformations that describe the act of changing reference frames form a group. Most discussion of quantum reference frames revolves around reference frames defined with respect to compact groups. For example, the relevant group used to describe a phase reference in quantum optics is $U(1)$ or the group used to describe the transformation between orientations  of a laboratory is $SO(3)$.

However, if we would like to apply the established formalism to more general groups, such as the Poincar\'{e} group and more generally to systems in curved spacetimes, we will need to understand quantum reference frames that are associated with non-compact groups. {The purpose of this paper is to embark on such an inquiry.}

We begin in Sec.~\ref{Relational descriptions} by introducing the $G$-twirl, which is a group average over all possible orientations of a system with respect to an external reference frame, and demonstrate its failure when naively applied to situations involving the non-compact groups of translations {in position and velocity}.  However, we find that the $G$-twirl over these groups naturally introduces a reduced state obtained by tracing out the center of mass degrees of freedom of a composite system. In Sec.~\ref{Relational encoding and the translation group} we examine informational properties of this reduced state for systems of two and three particles in fully separable Gaussian states with respect to an external frame. Specifically, we study the effective entanglement that ``appears'' when moving from a description of the system with respect to an external frame to a fully relational description,  which can alternatively be interpreted in terms of noise. This study is motivated by the need to determine how best to prepare states in the external partition in order to encode information in relational degrees of freedom, which will be useful for various communications tasks \cite{Checinska:2014}. We conclude in Sec.~\ref{Discussion} with a discussion and summary of the results presented.

\section{Relational descriptions}
\label{Relational descriptions}

In constructing a relational quantum theory, one essential task will be the description of a quantum system with respect to another quantum system. We thus seek a way in which to remove any information contained in a quantum state that makes reference to an external reference frame. This is accomplished by the $G$-twirl, which we introduce in Sec.~\ref{Relational description for compact groups} and apply to the group of translations and boosts\footnote{By boost it is meant Galilean boost, as opposed to Lorentz boost.} in Sec.~\ref{Relational description for non-compact groups}.

\subsection{Relational description for compact groups}
\label{Relational description for compact groups}

When the state of a system is described with respect to an external reference frame, such that the transformations that generate a change of this reference frame form a compact group, the relational description is well studied  \cite{Bartlett:2007}. 

Suppose we have a quantum system in the state $\rho \in \mathcal{B}(\mathcal{H})$, {where $\mathcal{B}(\mathcal{H})$ is the set of bounded linear operators on the Hilbert space $\mathcal{H}$}, described with respect to an external reference frame. Changes of the orientation of the system with respect to the external frame are generated by $U(g)$ acting on $\rho$, where $U(g)$ is the unitary representation of the group element $g\in G$, and $G$ is the compact group of all possible changes of the external reference frame. The relational description of $\rho$, that is the quantum state that does not contain any information about the external frame, is given by an average over all possible orientations of $\rho$ with respect to the external frame, with each possible orientation given an equal weight
\begin{align}
\mathcal{G} \! \left( \rho \right) := \int \dif{\mu\! \left(g\right)} \, U\!\left(g\right)\rho U^{\dagger}\!\left(g\right), \label{TwirlEncoding}
\end{align}
where $ \dif{\mu}(g)$ is the Haar measure of the group $G$; this averaging is referred to as the $G$-twirl. By averaging over all elements of the group, the $G$-twirl removes any relation to the external reference frame that was implicitly made use of in the description of $\rho$. What remains is only information about the relational degrees of freedom within the system. For example, if $\rho \in \mathcal{B}(\mathcal{H})$ describes a composite system of two particles such that $\mathcal{H}=\mathcal{H}_1 \otimes \mathcal{H}_2$, what remains in $\mathcal{G}(\rho)$ is information about the relational degrees of freedom between the two particles. Notice that the $G$-twirl is performed via the product representation $U(g)=U_1(g)\otimes U_2(g)$, where $U_1$ and $U_2$ are representations of the group $G$ for system $1$ and system $2$, respectively.

This relational description is used extensively in the study of quantum reference frames involving compact groups \cite{Bartlett:2007, Bartlett:2009, Jarzyna:2012, Marvian:2008, Palmer:2013}. However, when the $G$-twirl operation is generalized to the case where the group $G$ is non-compact, and thus does not admit a normalized Haar measure, it results in unnormalized states.

For example, let us consider the $G$-twirl of the state $\rho \in \mathcal{B}(\mathcal{H})$, where $\mathcal{H} \cong L_2 (\mathbb{R})$, over the non-compact group of spatial translations $T$ generated by the momentum operator $\hat{P}$. Expressing $\rho$ in the momentum basis we find
\begin{align}
\mathcal{G}_T\!\left(\rho \right) &= \int \dif{g} \,  e^{-i g \hat{P}}  \left(\int \dif{p} \dif{p'} \,  \rho\! \left(p, p' \right) \ket{p}\!\bra{p'} \right) e^{ig \hat{P}}  \nn \\
&= 2\pi \int \dif{p} \, \rho \!\left(p,p \right) \ket{p}\!\bra{p},
\end{align}
where $\dif{g}\!$ is the Haar measure associated with $T$ and in going from the first to the second line we have used the definition of the Dirac delta function $2 \pi \delta(p-p') = \int \dif g \, e^{ig(p-p')}$. Although the averaging operation is mathematically well defined, the resulting state $\mathcal{G}(\rho )$ is not normalized, as the trace of $\mathcal{G}_T(\rho )$ is infinite. This is a result of the Haar measure associated with $T$ not being normalized, i.e., the integral $\int \dif{g}\!$ is infinite. This issue does not arise when twirling over a compact group for which there exists a normalized Haar measure. Thus the relational description constructed by averaging a system over all possible orientations of a reference frame fails when the group describing changes of the reference frame is non-compact. 

One may try to remedy this problem by introducing a measure $p(g)$ on the group, such that $\int \dif g \, p(g) = 1$, and interpreting $p(g)$ as representing a priori knowledge of how the average should be performed \cite{Mehdi:2015}. However,  in general there is no objective way to choose $p(g)$---if we want a normalized measure it cannot be invariant.

\subsection{Relational description for non-compact groups}
\label{Relational description for non-compact groups}

We now construct a relational description of quantum states suitable for systems described with respect to reference frames associated with the non-compact groups of boosts and translations. We begin by twirling the state of a system of particles $\rho\in \mathcal{B}(\mathcal{H})$, over all possible boosts and translations of the external reference frame $\rho$ is specified with respect to. The result of this twirling is an unnormalized state proportional to $\mathbb{I}_{CM} \otimes \rho_R$, where $\mathbb{I}_{CM}$ is the identity on the center of mass degrees of freedom and $\rho_R=\tr_{CM} \rho$ is a normalized density matrix describing the relative degrees of freedom of the system. In doing so, we connect two approaches to quantum reference frames that have been studied in the past, specifically, the approach introduced by \citet{Bartlett:2007}, which makes use of the twirl to remove any information the state may have about an external reference frame, and the approach of \citet{Angelo:2011}, in which they trace over center of mass degrees of freedom to obtain a relational state.

Consider a composite system of $N$ particles each with mass $m_n$. We may partition the Hilbert space $\mathcal{H}$ of the entire system as $\mathcal{H} = \bigotimes_n \mathcal{H}_n$ where $\mathcal{H}_n \cong L_2(\mathbb{R}^3)$ which spans the degrees of freedom defined with respect to an external frame associated with the $n$th particle; we will refer to this as the external partition of the Hilbert space.  We may alternatively partition the Hilbert space as $\mathcal{H} = \mathcal{H}_{CM} \otimes \mathcal{H}_R$, where $\mathcal{H}_{CM} \cong L_2(\mathbb{R}^3)$ is associated with the degrees of freedom of the centre of mass defined with respect to an external frame, and $\mathcal{H}_R\cong L_2(\mathbb{R}^{3N-3})$ is associated with the relative degrees of freedom of the system defined with respect to a chosen reference particle; we will refer to this partition as the center of mass and relational partition of the Hilbert space.

As was done in Sec.~\ref{Relational description for compact groups} for reference frames associated with compact groups, to obtain a relational state we will average the state of our system over all possible orientations---intended in a generic sense, meant here to be about translations and boosts---with respect to the external frame. Here we consider the system to be described with respect to an inertial external frame. Thus a change of the external frame corresponds to acting on the system with an element of the Galilean group, and the average  over all possible orientations of the system with respect to the external frame will be an average over the Galilean group. 

The Galilean group, $Gal$, is a semidirect product of the translation group $T_4$, the group of boosts \enlargethispage{200pt} 
$B_3$, and the rotation group $SO(3)$:
\begin{align}
Gal \cong T_4 \rtimes \Big( B_3 \rtimes SO(3) \Big).
\end{align}
We will restrict our analysis to an average over spatial translations $T_3$, where $T_4 \cong T_1 \rtimes T_3$, and boosts $B_3$, as averages over $SO(3)$, the orientation of a system with respect to an external frame, have been well studied in literature \cite{Bartlett:2007}, and we are primarily interested in issues associated with non-compact groups. Further, we do not average over time translations $T_1$ as this would require us to introduce a Hamiltonian to generate time translations, and for now we are interested only in a relative description of the state at one instant of time and not its dynamics. Suppose the state of a system was given with respect to an external reference frame with a specific position and velocity. The operator that results from these restricted averages is related to the state as seen from an observer who is ignorant of both the position and velocity of the external reference frame.

The position and momentum operators associated with the centre of mass, $\hat{\mathbf{X}}_{CM}$ and $\hat{\mathbf{P}}_{CM}$, and relational degrees of freedom, $\hat{\mathbf{X}}_{i|1}$ and $\hat{\mathbf{P}}_{i|1}$, may be expressed in terms of the operators $\hat{\mathbf{X}}_n$ and $\hat{\mathbf{P}}_n$ associated with the position and momentum operators of each of the $N$ particles with respect to the external frame as
\begin{subequations}
\label{CMRcoordinates}
\begin{align}
\hat{\mathbf{X}}_{CM} &= \frac{1}{\sum_n m_n} \sum_n m_n \hat{\mathbf{X}}_n ,\\
\hat{\mathbf{P}}_{CM} &= \sum_n P_n, \\
\hat{\mathbf{X}}_{i|1} &= \hat{\mathbf{X}}_i - \hat{\mathbf{X}}_1  \ {\rm for} \ i \in \{2, N\}, 
\end{align}
\end{subequations}
and the relative momentum operators $\hat{\mathbf{P}}_{i|1}$ are chosen such that they satisfy the canonical commutation relations $[ \hat{\mathbf{X}}_{i|1},\hat{\mathbf{P}}_{j|1}] =i \delta_{ij}$ and all other commentators vanish\footnote{This choice of operators on $\mathcal{H}_R$ is not unique. We may have alternatively defined a set of $N-1$ relative momentum operators and defined the $N-1$ relative position operators as those which satisfy the canonical commutation relations. See \cite{Angelo:2011} for more details.}. Without loss of generality we have chosen to define the relative degrees of freedom with respect to particle 1.

The action of a translation $\mathbf{g} \in \mathbb{R}^3 \cong  T_3$ and boost $\mathbf{h} \in \mathbb{R}^3 \cong B_3$ of the external frame in the external partition $\mathcal{H} = \bigotimes_n \mathcal{H}_n$  is given by
\begin{subequations}
\begin{align}
U_T(\mathbf{g}) &= \bigotimes_n e^{-i\mathbf{g} \cdot \hat{\mathbf{P}}_n}, \\
U_B(\mathbf{h}) &= \bigotimes_n e^{i m_n  \mathbf{h} \cdot \hat{\mathbf{X}}_n },
\end{align}
\end{subequations}
and in the center of mass and relational partition $\mathcal{H}_{CM} \otimes \mathcal{H}_R$ is given by
\begin{subequations}
\begin{align}
U_T(\mathbf{g}) &=  e^{-i\mathbf{g} \cdot \hat{\mathbf{P}}_{CM} } \otimes \mathbb{I}_R, \label{transU} \\
U_B(\mathbf{h}) &=  e^{iM \mathbf{h} \cdot \hat{\mathbf{X}}_{CM}}  \otimes \mathbb{I}_R, \label{boostU}
\end{align}
\label{TranslationsBoostsGlobal}
\end{subequations}
where  $M = \sum_n m_n$ is the total mass.

To carry out the average over $T_3$ and $B_3$, let us express $\rho$ in the $ \mathcal{H}_{CM} \otimes \mathcal{H}_R$ partition in the momentum basis
\begin{align}
\rho &= \int  \dif{\mathbf{p}_{CM}}  \dif{\mathbf{p}_{CM}'}   \dif{\mathbf{p}_R}  \dif{\mathbf{p}_R'} \, \rho \! \left(\mathbf{p}_{CM},\mathbf{p}_{CM}', \mathbf{p}_R,\mathbf{p}_R' \right) \nn \\
  & \qquad \ket{\mathbf{p}_{CM}}\!\bra{\mathbf{p}_{CM}'} \otimes \ket{\mathbf{p}_R} \!\bra{\mathbf{p}_R'},
\end{align}
where $\mathbf{p}_{CM}$ and $\mathbf{p}_{CM}'$ denote the momentum vector of the center of mass and $\mathbf{p}_R$ and $\mathbf{p}'_R$ denote the $N-1$ relative momentum vectors. Making use of Eq.~\eqref{transU}, we may average over all possible spatial translations of the external frame
\begin{widetext}
\begin{align}
\mathcal{G}_T \! \left(\rho\right) &=   \int  \dif{\mathbf{p}_{CM}} \dif{\mathbf{p}_{CM}'}   \dif{\mathbf{p}_R}   \dif \mathbf{p}_R'  \, \rho \!\left(\mathbf{p}_{CM},\mathbf{p}_{CM}', \mathbf{p}_R,\mathbf{p}_R' \right)  \int \dif{\mathbf{g}} \, U_T(\mathbf{g})\ket{\mathbf{p}_{CM}}\!\bra{\mathbf{p}_{CM}'} U_T(\mathbf{g})^{\dagger} \otimes \ket{\mathbf{p}_R} \! \bra{\mathbf{p}_R'} \nn \\
&=   2 \pi \int   \dif \mathbf{p}_{CM}   \dif \mathbf{p}_R  \dif \mathbf{p}_R'  \, \rho \! \left(\mathbf{p}_{CM},\mathbf{p}_{CM}, \mathbf{p}_R,\mathbf{p}_R' \right) \ket{\mathbf{p}_{CM}}\!\bra{\mathbf{p}_{CM}}  \otimes \ket{\mathbf{p}_R}\! \bra{\mathbf{p}_R'}. \label{translationAverage}
\end{align}
The effect of averaging over all possible translations is to project $\rho$ into a charge sector of definite center of mass momentum. Now averaging Eq. \eqref{translationAverage} over all boosts, using Eq. \eqref{boostU}, yields
\begin{align}
\mathcal{G}_B \circ \mathcal{G}_T \! \left(\rho\right) &= 2 \pi \int \dif \mathbf{h}   \int  \dif \mathbf{p}_{CM}   \dif \mathbf{p}_R  \dif \mathbf{p}_R'  \, \rho \! \left(\mathbf{p}_{CM},\mathbf{p}_{CM}, \mathbf{p}_R, \mathbf{p}_R' \right)   U_B(\mathbf{h})  \ket{\mathbf{p}_{CM}}\!\bra{\mathbf{p}_{CM}}U_B(\mathbf{h})^{\dagger}  \otimes \ket{\mathbf{p}_R} \!\bra{\mathbf{p}_R'} \nn \\
&= 2 \pi\int \dif \mathbf{h}   \int  \dif p_{CM}  \dif \mathbf{p}_R  \dif \mathbf{p}_R'  \, \rho \! \left(\mathbf{p}_{CM}-M\mathbf{h},\mathbf{p}_{CM}-M\mathbf{h}, \mathbf{p}_R, \mathbf{p}_R' \right)  \ket{\mathbf{p}_{CM} }\!\bra{\mathbf{p}_{CM}}  \otimes \ket{\mathbf{p}_R}\! \bra{\mathbf{p}_R'} \nn \\
&= \frac{2 \pi}{M}\int \dif \mathbf{h}   \int  \dif \mathbf{p}_{CM}  \dif \mathbf{p}_{R}  \dif \mathbf{p}_R'  \, \rho \! \left(\mathbf{h}, \mathbf{h}, \mathbf{p}_R, \mathbf{p}_R' \right)  \ket{\mathbf{p}_{CM} }\!\bra{\mathbf{p}_{CM}}  \otimes \ket{\mathbf{p}_R} \!\bra{\mathbf{p}_R'} \nn \\
&=  \frac{2 \pi}{M}  \int  \dif \mathbf{p}_{CM}  \ket{\mathbf{p}_{CM} }\!\bra{\mathbf{p}_{CM}}  \otimes \int \dif \mathbf{p}_R  \dif \mathbf{p}_R'  \,  \left(\int \dif \mathbf{h} \, \rho \! \left(\mathbf{h}, \mathbf{h}, \mathbf{p}_R, \mathbf{p}_R' \right)\right)  \ket{\mathbf{p}_R} \!\bra{\mathbf{p}_R'} \nn \\
&=   \frac{2 \pi}{M} \mathbb{I}_{CM}  \otimes \rho_R, \label{BoostTranslationAverage}
\end{align}
\end{widetext}
where in the last line
\begin{align}
\rho_R =&  \tr_{CM} \rho \nn \\
=& \int \dif \mathbf{p}_R  \dif \mathbf{p}_R'  \,  \left(\int \dif \mathbf{h} \, \rho \! \left(\mathbf{h}, \mathbf{h}, \mathbf{p}_R, \mathbf{p}_R' \right)\right)  \ket{\mathbf{p}_R} \!\bra{\mathbf{p}_R'} \label{RelationEncoding},
\end{align}
and we have made use of the resolution of the identity $\mathbb{I}_{CM} =  \int  \dif \mathbf{p}_{CM} \, \ket{\mathbf{p}_{CM} }\!\bra{\mathbf{p}_{CM}}$. 

From the appearance of the identity $\mathbb{I}_{CM}$ in Eq.~\eqref{BoostTranslationAverage}, we see that $\mathcal{G}_B \circ \mathcal{G}_T (\rho)$ contains no information about the center of mass, and thus no information about the external frame. As discussed earlier, since we have averaged over a non-compact group, Eq.~\eqref{BoostTranslationAverage} is unnormalizable, and thus $\mathcal{G}_B \circ \mathcal{G}_T (\rho)$ is not a physical state. However, all the information about the relational degrees of freedom of the system is encoded in $\rho_R$, which is normalized.

By twirling over all possible boosts and translations of the system, we see from Eq.~\eqref{BoostTranslationAverage} that the reduced state $\rho_R$ naturally appears. We have thus connected the use of $\rho_R$ that is made in \citet{Angelo:2011} when analyzing absolute and relative degrees of freedom, with the usual quantum reference formalism \cite{Bartlett:2007}.

In general, when transforming from the external partition $\mathcal{H} = \bigotimes_n \mathcal{H}_n$, to the center of mass and relational partition $\mathcal{H} = \mathcal{H}_{CM} \otimes \mathcal{H}_R$, entanglement will appear between the center of mass and relational degrees of freedom, as well as within the relational Hilbert space $\mathcal{H}_R$. Thus the state $\rho_R$ will be mixed, reflecting the fact that information about the external degrees of freedom has been lost. This is analogous to information about the external frame being lost in Eq. \eqref{TwirlEncoding} when averaging over all elements of a compact group.

\section{Gaussian quantum mechanics and the relational description}
\label{Relational encoding and the translation group}

We now examine in detail the informational properties of the  reduced state $\rho_R$ of the relational degrees of freedom given in Eq. \eqref{RelationEncoding}, by examining systems of two and three particles in one dimension distinguished by their masses. As mentioned earlier, in general, entanglement will appear when moving from the external partition $\mathcal{H} = \bigotimes_n \mathcal{H}_n$, to the center of mass and relational partition $\mathcal{H} = \mathcal{H}_{CM} \otimes \mathcal{H}_R$. This entanglement is crucial in determining how to describe physics relative to a particle within the system~\cite{Angelo:2011}. For example, if there is entanglement between the centre of mass and the relational degrees of freedom, an observer identified with the reference particle, particle 1 as chosen in Eq.~\eqref{CMRcoordinates}, will describe the rest of the system as being in a mixed state.

As a concrete example of the entanglement that can emerge when changing from the external partition to the center of mass and relational partition of the Hilbert space, we consider systems of two and three particles in Gaussian states in the external partition. The advantage of considering Gaussian states in the external partition is that the transformation which takes the state from being specified in the external partition to being specified in the centre of mass and relational partition is a Gaussian unitary, that is, a state which is Gaussian in the external partition will also  be Gaussian in the center of mass and relational partition. Further, if we are interested in the reduced state $\rho_R$ defined in Eq.~\eqref{RelationEncoding}, and the state of the particles in either partition is a Gaussian state, then the trace over the centre of mass degrees of freedom also results in a Gaussian state. Thus, by considering Gaussian states in the external partition we are able to make use of the extensive tools developed in the field of Gaussian quantum information. We begin here by briefly reviewing relevant aspects of Gaussian quantum information; for more detail the reader may consult one of the many good references on the topic \cite{Adesso:2007, Quantum-Information:2011, Adesso:2014}.

\subsection{The Wigner function and Gaussian states}
\label{The Wigner function}

Any density operator has an equivalent representation as a quasi-probability distribution over phase space. To see this, we introduce the Weyl operator
\begin{align}
D\!\left(\boldsymbol{\xi}\right) := \exp \! \left( i \hat{\mathbf{x}}^T \boldsymbol{\Omega} \boldsymbol{\xi} \right),
\end{align}
where $\hat{\mathbf{x}} := (\hat{q}_1,\hat{p}_1,\ldots, \hat{q}_n,\hat{p}_n)$ is a vector of phase space operators, $\boldsymbol{\xi} \in \mathbb{R}^{2n}$, and $\Omega$ is the symplectic form defined as 
\begin{align}
\Omega=\bigoplus_{i=1}^n \omega, \quad \mbox{with} \quad  \omega = 
\begin{pmatrix} 
0 & 1\\
-1 & 0
\end{pmatrix}.
\end{align}
A density operator $\rho \in \mathcal{B} (\mathcal{H})$ has an equivalent representation as a Wigner characteristic function $\chi(\boldsymbol{\xi}) := \tr [ \rho D (\boldsymbol{\xi}) ]$, or by its Fourier transform, known as the Wigner function
\begin{align}
W\left( \mathbf{x} \right) := \int_{\mathbb{R}^{2n}} \frac{\dif^{2n }\xi}{{\left(2 \pi\right)^{2n}}} \exp \! \left( - i \mathbf{x}^T \boldsymbol{\Omega} \boldsymbol{\xi} \right) \chi \left( \boldsymbol{\xi}\right). \label{WignerDef}
\end{align}
where $\mathbf{x} := (q_1,p_1,\dots,q_n,p_n)$ is a vector of phase space variables.

An $n$-particle Gaussian state  is a state whose Wigner function is Gaussian, that is
\begin{align}
W\left(\mathbf{x}; \bar{\mathbf{x}}, \mathbf{V} \right) = \frac{ \exp \! \left({-\frac{1}{2} \left( \mathbf{x} - \bar{\mathbf{x}}\right)^T \mathbf{V}^{-1}  \left( \mathbf{x} - \bar{\mathbf{x}}\right) }\right)}{\left(2 \pi\right)^n \sqrt{\det \mathbf{V}}}, \label{GaussianState}
\end{align}
where $\bar{\mathbf{x}} := (\bar{q}_1,\bar{p}_1,\dots,\bar{q}_n,\bar{p}_n)$ is given by a vector of averages 
\begin{align}
\bar{x}_i := \braket{\hat{x}_i} = \tr \left[ \hat{x}_i \rho \right],
\end{align}
and $\mathbf{V}$ is the real $2n \times 2n$ covariance matrix with components
\begin{align}
V_{ij} := \frac{1}{2} \tr \left[ \left\{\hat{x}_i - \bar{x}_i, \hat{x}_j - \bar{x}_j\right\} \rho \right],
\end{align}
where we have made use of the anticommutator $\left\{ A, B \right\} := AB + BA$.

\subsection{Two particles}
\label{Two particles}

We begin our analysis by considering two particles with masses $m_1$ and $m_2$ to be in a tensor product of Gaussian states $\rho_E = \rho_1 \otimes \rho_2$, where $\rho_1 \in \mathcal{B}(\mathcal{H}_1)$ and $\rho_2\in \mathcal{B}(\mathcal{H}_2)$ in the external partition $\mathcal{H} = \mathcal{H}_1 \otimes \mathcal{H}_2$. Due to the tensor product structure of $\rho_E$, the Wigner function of the composite system is a product of the Wigner functions associated with particles 1 and 2
\begin{align}
W\left(\mathbf{x}; \bar{\mathbf{x}}_E, \mathbf{V}_E \right) = W\left(\mathbf{x}; \bar{\mathbf{x}}_1, \mathbf{V}_1 \right)  W\left(\mathbf{x}; \bar{\mathbf{x}}_2, \mathbf{V}_2 \right).
\end{align}

The reason for considering factorized states in the external partition, apart from their common usage in the literature \cite{Palmer:2013,Bartlett:2009}, is that if we are to use the composite system for communication, the tensor product structure is easily prepared as it does not require an entangling operation. Further, if one party wishes to communicate a string of classical bits (or qubits), they can try to encode one bit (or qubit) per physical qubit, and this string can be decoded sequentially. The sender does not need to know at the outset the entire message they wish to communicate, and the receiver does not need to store the entire message before decoding it \cite{Bartlett:2009}. 

As we will only be interested in the entanglement generated in moving from the external partition to the center of mass and relational partition, we may, without loss of generality, set $\bar{\mathbf{x}}_1= \bar{\mathbf{x}}_2 = 0$ as these averages can be arbitrarily adjusted via local unitary operations in either partition, and thus do not affect the entanglement properties under consideration. 

Making use of Eq. \eqref{GaussianState}, we find the covariance matrix associated with $\rho_E$ is given by $\mathbf{V}_{E} = \mathbf{V}_{1} \oplus \mathbf{V}_{2}$; the direct sum structure resulting from the fact the we chose $\rho_E$ to be a tensor product state in the external partition. Using Williamson's theorem \cite{Williamson:1936}, one can show that the most general form of the covariance matrices $\mathbf{V}_1$ and $\mathbf{V}_2$ is given by
\begin{align}
\mathbf{V}_{i} &= \frac{1}{\mu_i} \mathbf{R}\left(\theta_i\right) \mathbf{S} \left(2r_i\right) \mathbf{R}\left(\theta_i\right)^T \nn \\
& = \frac{1}{\mu_i}
\begin{psmallmatrix}
\cosh 2r_i - \cos 2 \theta_i \sinh 2r_i & \sin 2 \theta_i \sinh 2r_i \\
\sin 2 \theta_i \sinh 2r_i & \cosh 2r_i + \cos 2 \theta_i \sinh 2r_i
\end{psmallmatrix}, \label{singlemodecovaraince}
\end{align}
where  the free parameter $\mu_i = 1/\sqrt{\det \mathbf{V}_i}  \in (0,1]$ is the purity, $\tr(\rho_i^2)$, of the state $\rho_i$, $\mathbf{R}\left(\theta_i\right)$ is a rotation matrix specifying a phase rotation by an angle $\theta_i \in [0,\pi/4]$, and $\mathbf{S}(2r_i)$ is a diagonal symplectic matrix specifying a squeezing of the Wigner function parameterized by $ r_i \in \mathbb{R}$.

\subsubsection{Transforming to the center of mass and relational partition}
\label{Transforming to the global and relational partition}

For two particles in one dimension the transformation from the external degrees of freedom $\mathbf{x}_E := (x_1, p_1, x_2, p_2)$, where $x_i$ and $p_i$ denote the position and momentum of the $i$th particle with respect to an external frame, to the center of mass and relational degrees of freedom $\mathbf{x}_{CMR} := (x_{cm}, p_{cm}, x_{2|1}, p_{2|1} )$, where $x_{cm}, p_{cm}$ are the position and momentum of the center of mass with respect to an external frame and $x_{2|1}, p_{2|1}$ are the position and momentum of particle~2 with respect to particle~1, is given by Eq.~\eqref{CMRcoordinates} with $N=2$ and vectors of operators replaced by a single operator. Under this transformation the external covariance matrix $\mathbf{V}_{E}$ transforms to $\mathbf{V}_{CMR} = \mathbf{M}_2 \mathbf{V}_{E} \mathbf{M}_2^T$, where $\mathbf{M}_2$ is given by
\begin{align}
\mathbf{M}_2 :=
\begin{pmatrix}
\frac{m_1}{m_1+m_2} & 0 & \frac{m_2}{m_1+m_2}& 0 \\
0 & 1 & 0 & 1 \\
-1 & 0 & 1 & 0\\
0 & -\frac{m_2}{m_1+m_2} & 0& 1- \frac{m_2}{m_1+m_2}
\end{pmatrix}. \label{2particleM}
\end{align}

As both the external and center of mass and relational position and momentum operators obey the canonical commutation relations, it follows that $\mathbf{M}_2$ is a symplectic transformation, i.e. it preserves the symplectic form $\mathbf{M}_2 \boldsymbol{\Omega} \mathbf{M}_2^T = \boldsymbol{\Omega}$. Since $\mathbf{M}_2$ is symplectic, the associated transformation preserves the Gaussianity of the state, that is, if a state is Gaussian in the external partition, it will also be Gaussian in the center of mass and relational partition. 

The relational state $\rho_R$ given in Eq. \eqref{RelationEncoding}, is a Gaussian state whose covariance matrix $\mathbf{V}_{2|1}$  is obtained by deleting the first and second rows and columns of $\mathbf{V}_{CMR}$; taking the most general form of $\mathbf{V}_1$ and $\mathbf{V}_2$ yields
\begin{align}
\mathbf{V}_{2|1} = \frac{1}{\mu_1 \mu_2}
\begin{psmallmatrix}
 \mu_2 f^{-}_1 +  \mu_1 f^{-}_2 & - \mu_2 \tilde{m}_2 g_1 +  \mu_1 \tilde{m}_1 g_2 \\
 - \mu_2 \tilde{m}_2 g_1 +  \mu_1 \tilde{m}_1 g_2 &  \mu_2 \tilde{m}_2 ^2 f^{+}_1 +  \mu_1 \tilde{m}_1^2  f^{+}_2
\end{psmallmatrix}, \label{RelationalState21}
\end{align}
where
\begin{align}
f^{\pm}_i &:= \cosh 2r_i \pm \cos 2 \theta_i \sinh 2r_i \nn, \\
g_i &:= \sin 2 \theta_i \sinh 2r_i, \nn
\end{align}
and $\tilde{m}_i: = m_i/(m_1+m_2)$.

\subsubsection{Entanglement between the center of mass and relational degrees of freedom}
\label{Entanglement between the global and relational degrees of freedom} 

\begin{figure}
\subfloat[$\theta = 0$]{%
\includegraphics[width=0.25\textwidth]{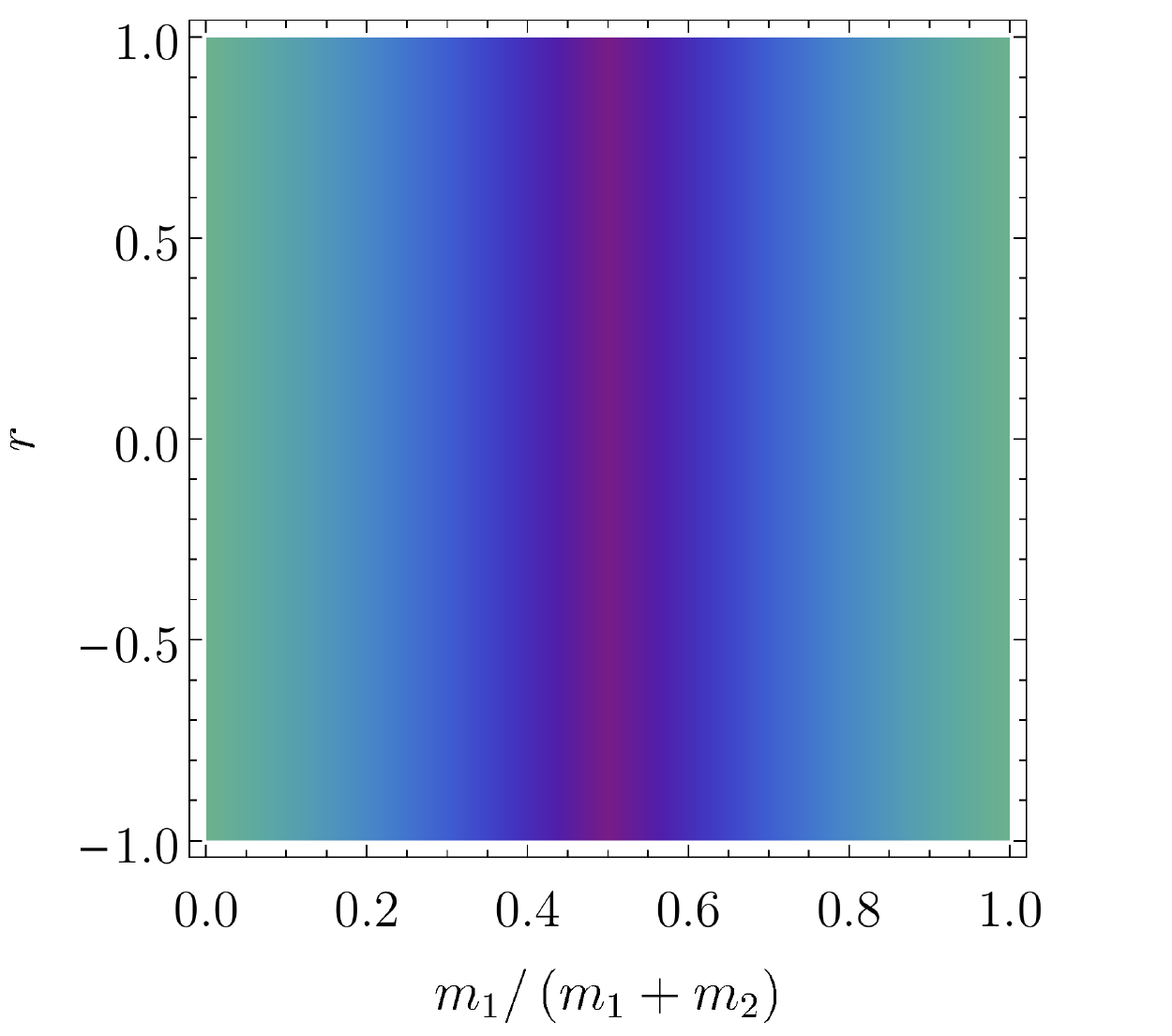}
\label{1a}
}
\subfloat[$\theta = \pi/32$]{%
\includegraphics[width=0.25\textwidth]{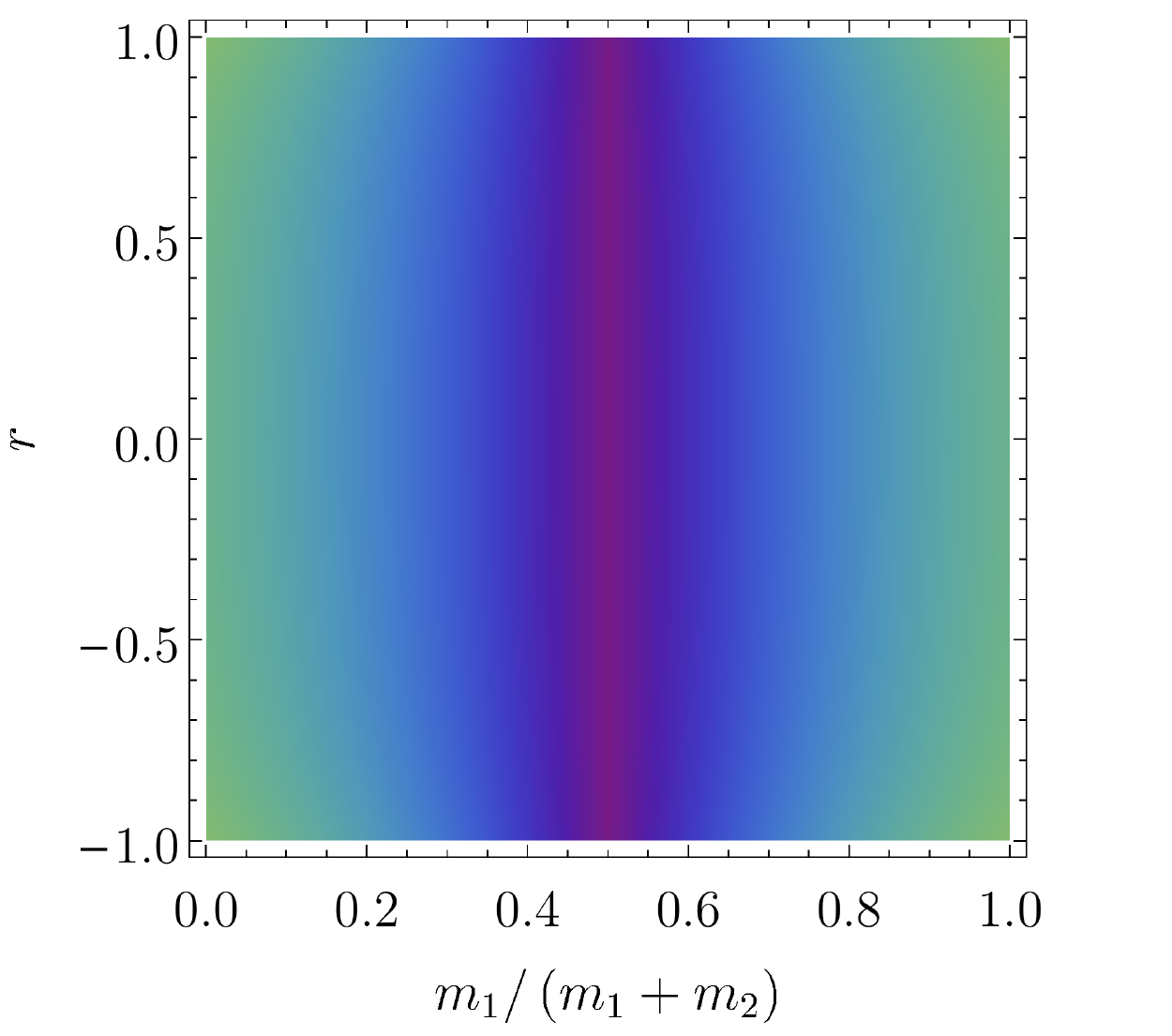}
\label{1b}
}
\\
\subfloat[$\theta = \pi/8$]{%
\includegraphics[width=0.25\textwidth]{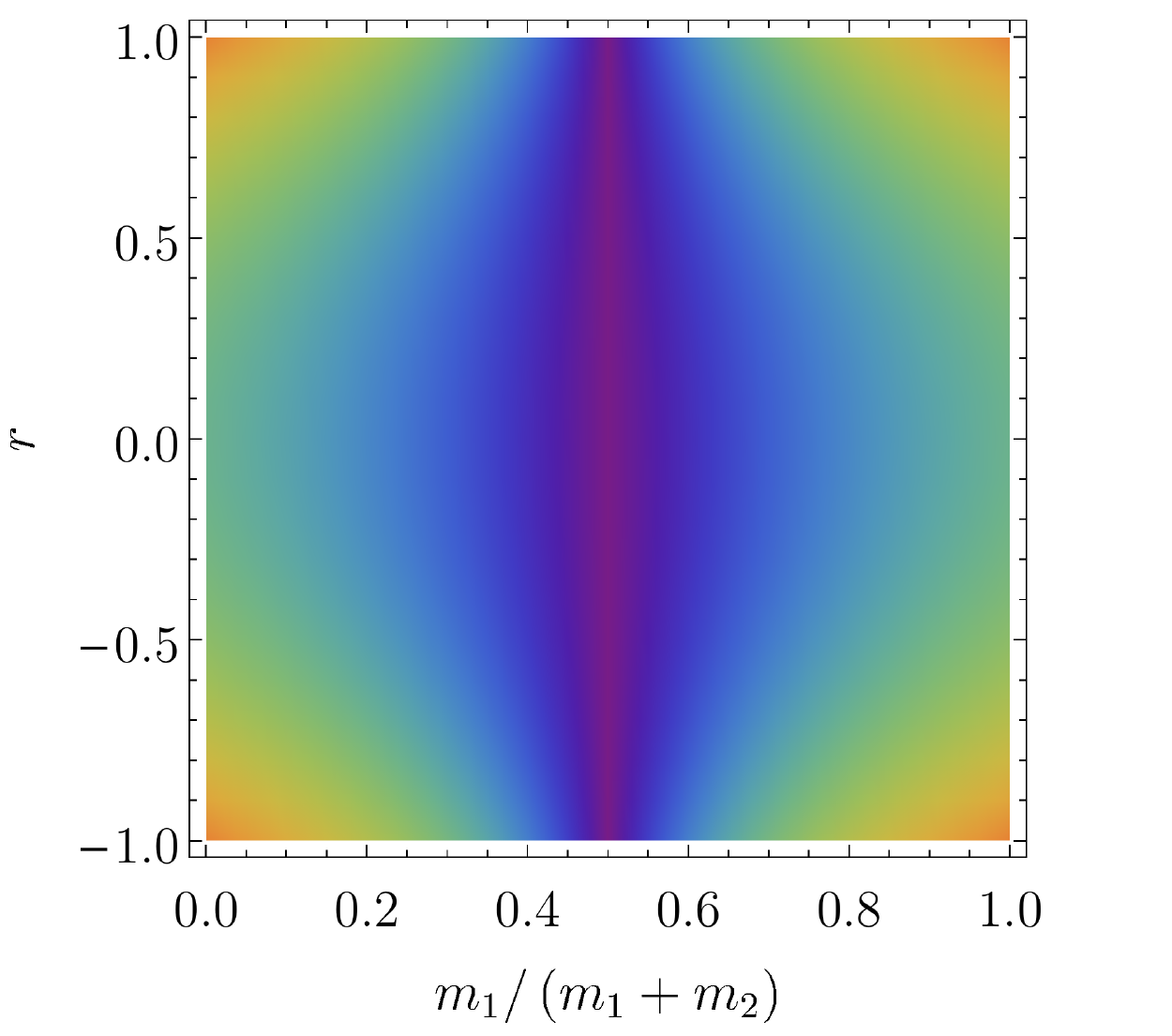}
\label{1c}
}
\subfloat[$\theta = \pi/4$]{%
\includegraphics[width=0.25\textwidth]{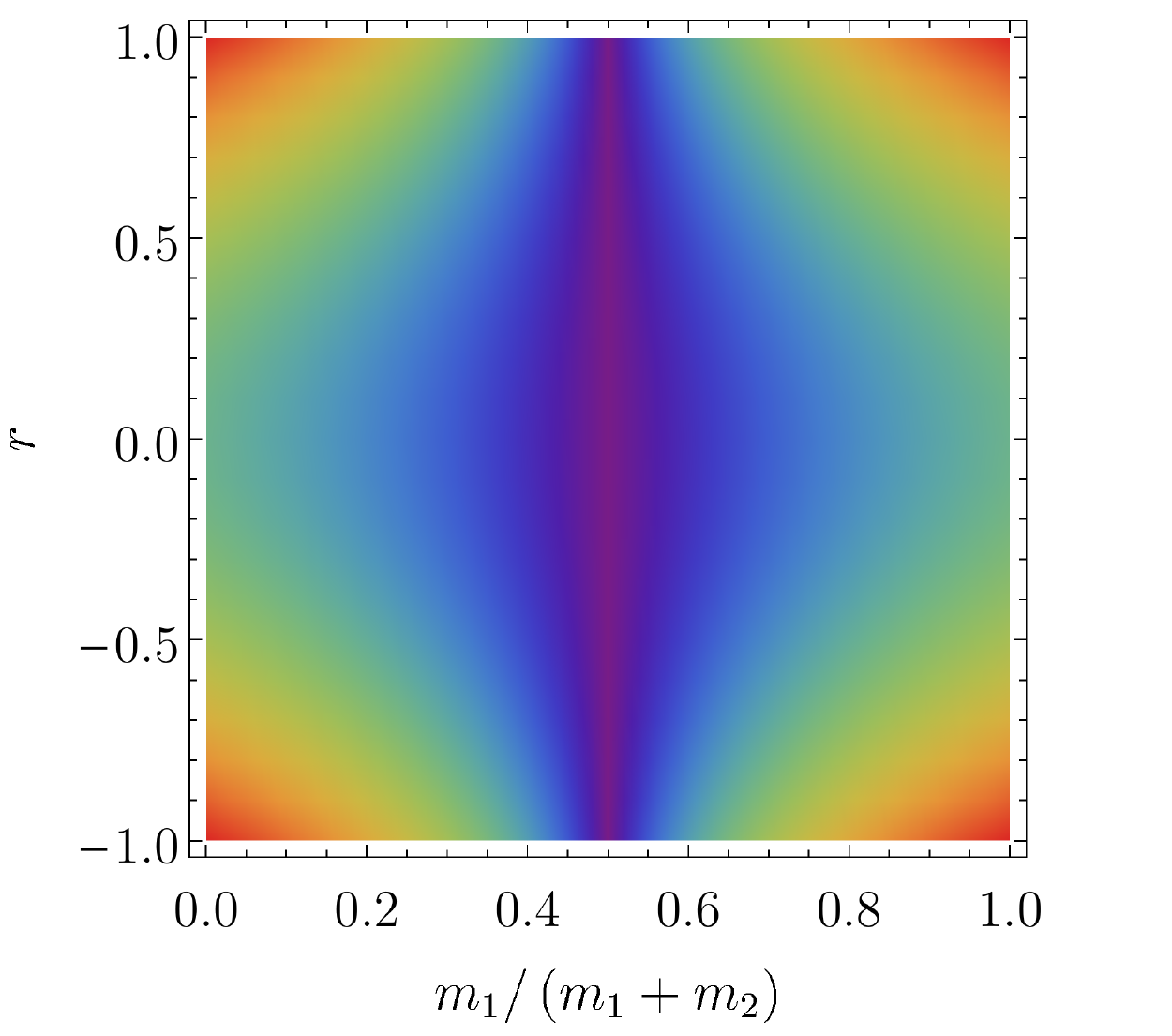}
\label{1d}
}
\\
\subfloat{%
\includegraphics[width=0.25\textwidth]{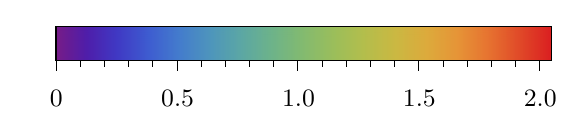}
}
\caption{(Colour online) The logarithmic negativity, as a measure of the entanglement between the center of mass and relation degrees of freedom, of the state associated with $\mathbf{V}_{CMR}$, when $\mathbf{V}_1 = \mathbf{V}_2$ and both $\rho_1$ and $\rho_2$ are pure, i.e. $\det \mathbf{V}_1 = \det \mathbf{V}_2 =1$, for different phase rotations $\theta=\theta_1=\theta_2$ as a function of the squeezing parameter $r=r_1=r_2$ and the ratio of masses $m_1/(m_1+m_2)$.}
\label{fig:identicalV}
\end{figure}


\begin{figure}
\subfloat[$\theta = 0$ and $\mu_1 = 0.6$]{%
\includegraphics[width=0.25\textwidth]{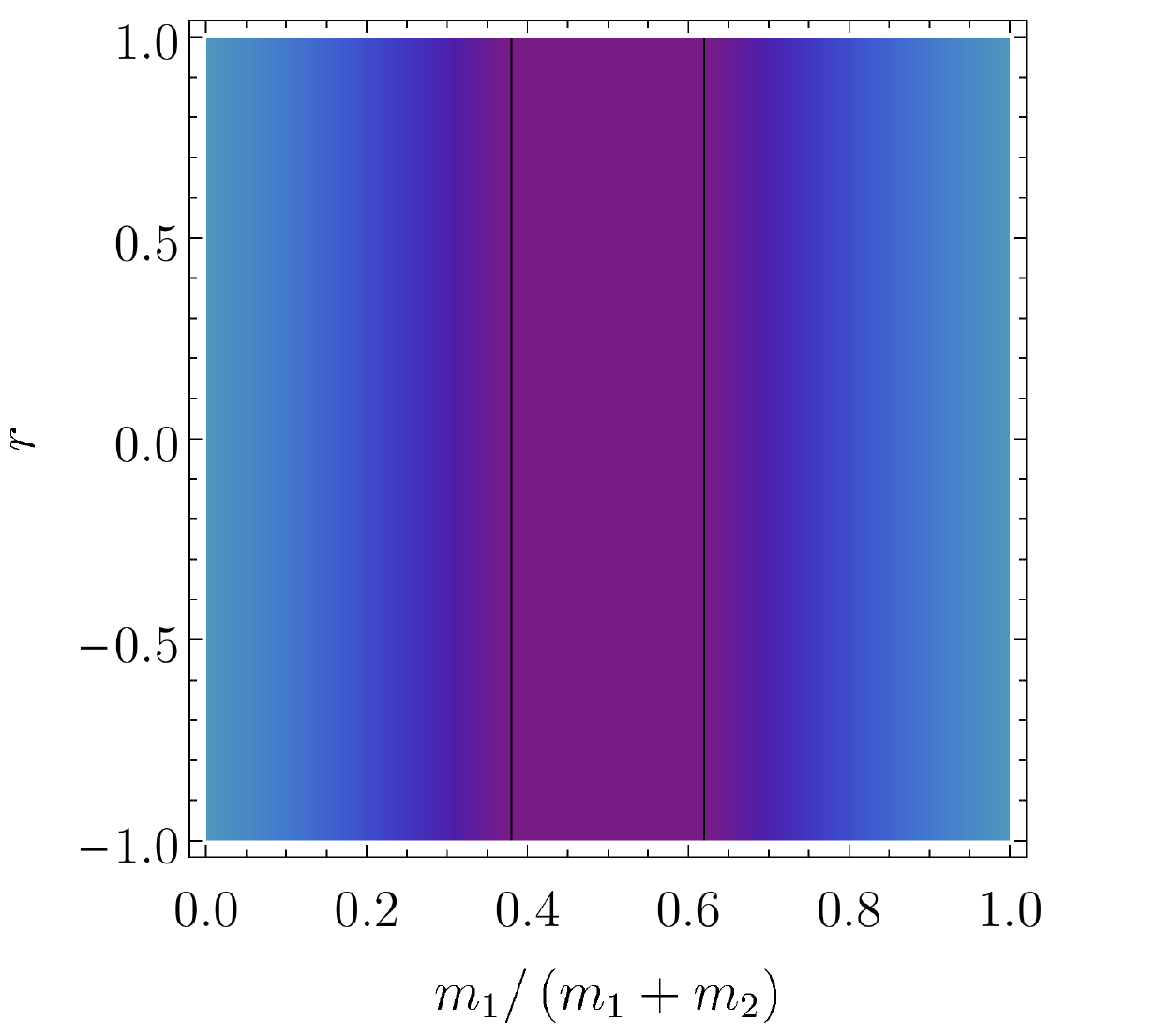}
}
\subfloat[$\theta = 0$ and $\mu_1 = 0.2$]{%
\includegraphics[width=0.25\textwidth]{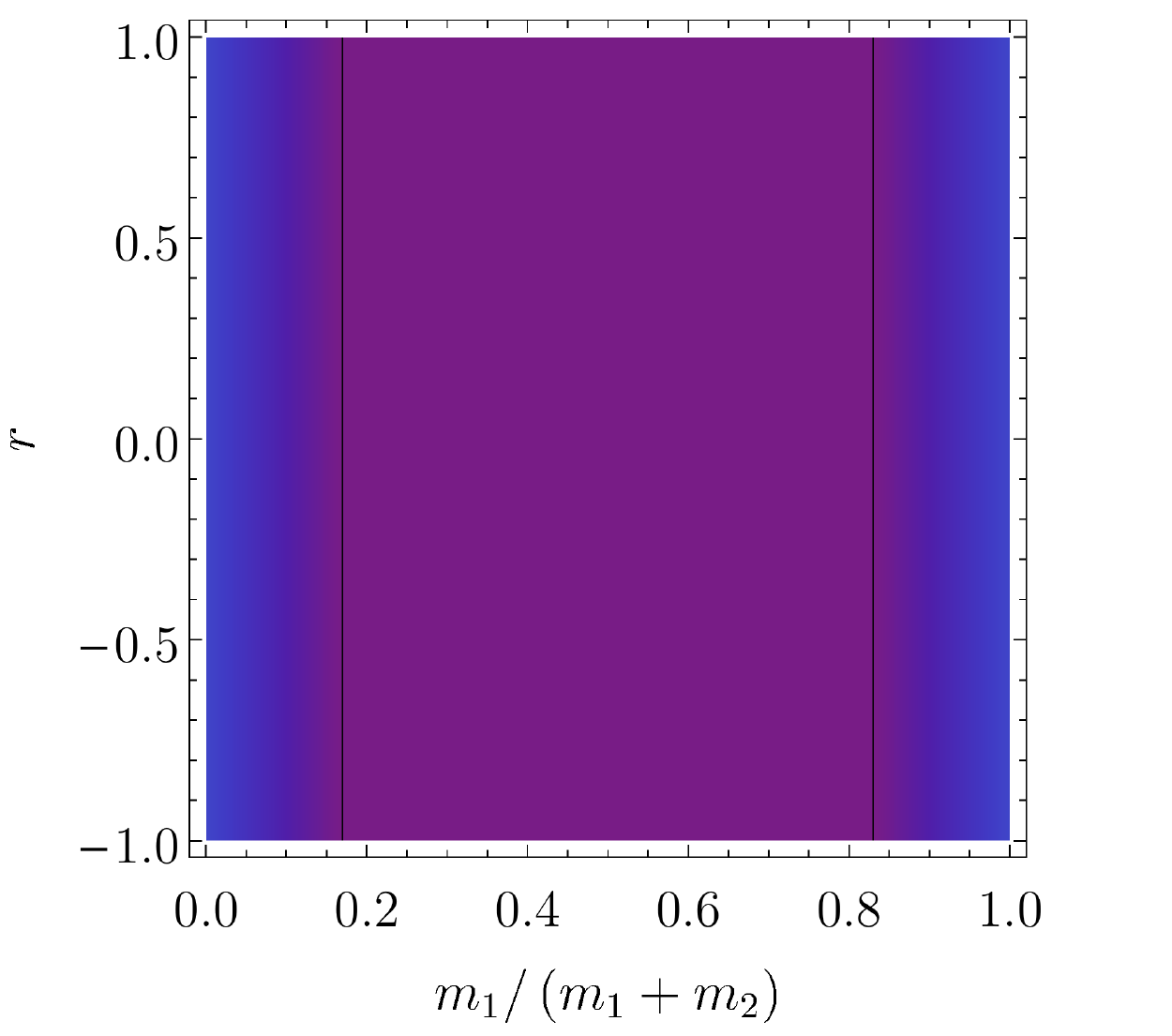}
}
\\
\subfloat[$\theta = \pi/4$ and $\mu_1 = 0.6$]{%
\includegraphics[width=0.25\textwidth]{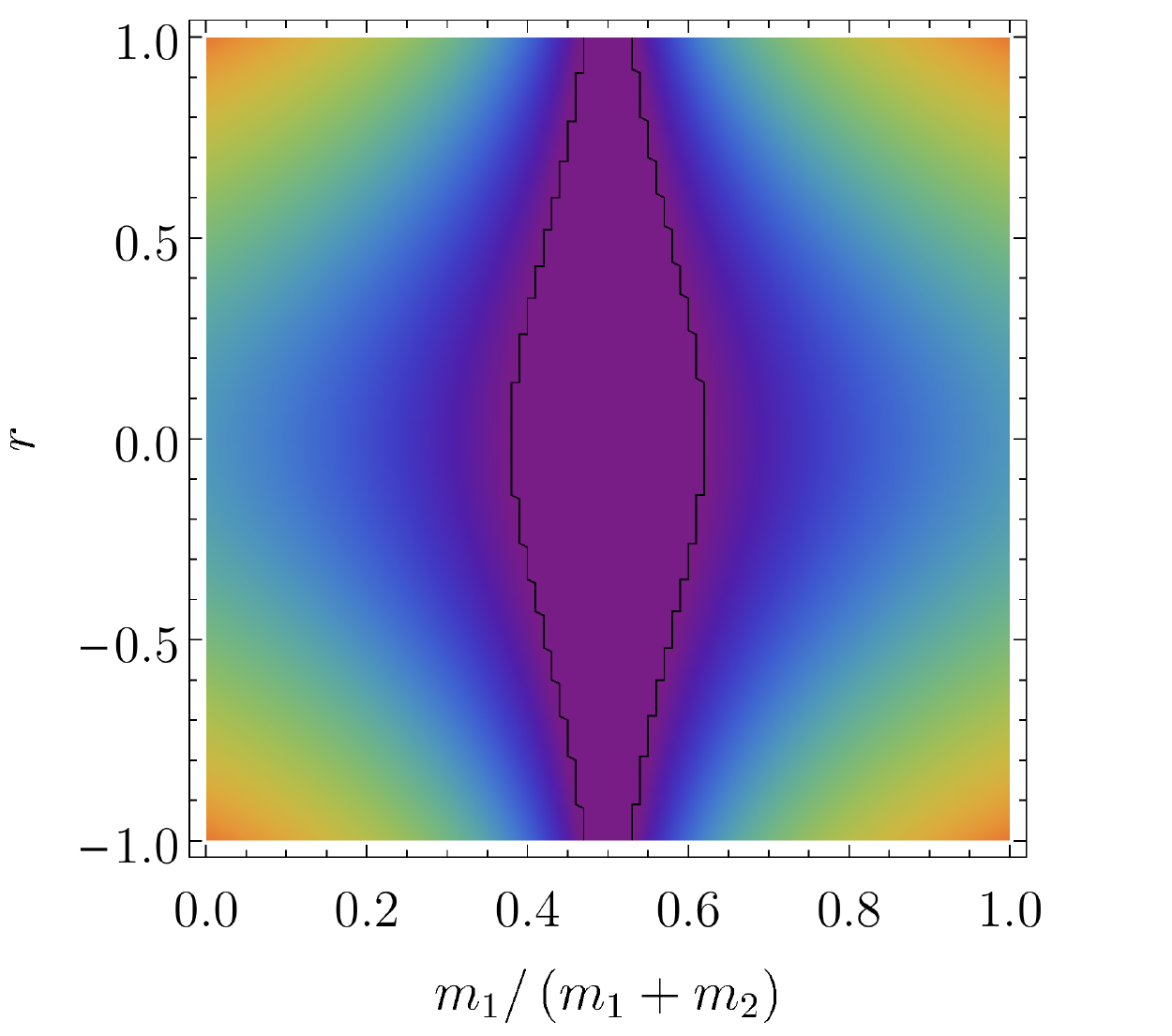}
}
\subfloat[$\theta = \pi/4$ and $\mu_1 = 0.2$]{%
\includegraphics[width=0.25\textwidth]{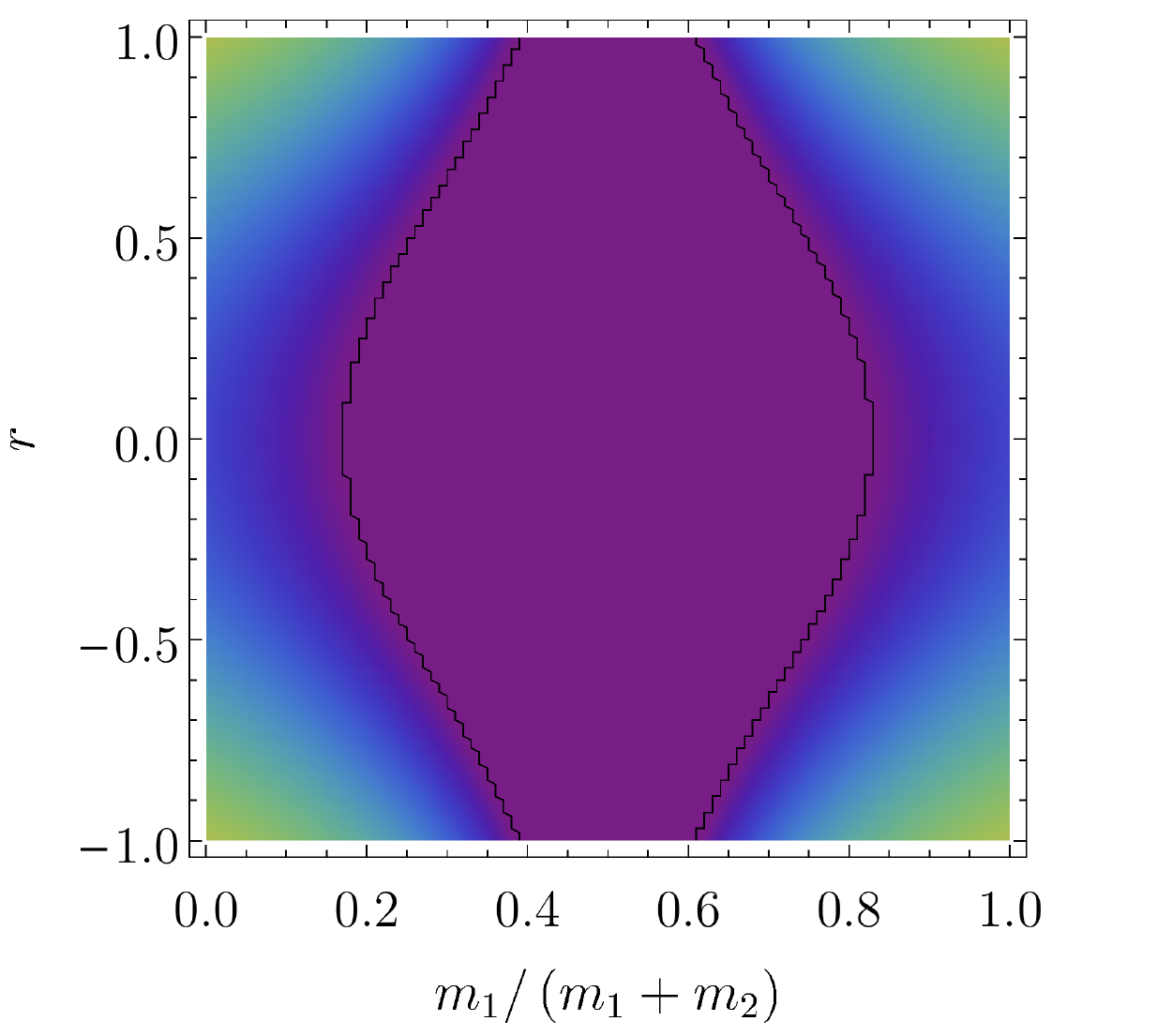}
}
\\
\subfloat{%
\includegraphics[width=0.25\textwidth]{legend.pdf}
}
\caption{(Colour online) The logarithmic negativity, as a measure of the entanglement between the center of mass and relation degrees of freedom, of the state associated with $\mathbf{V}_{CMR}$, when $r=r_1=r_2$, $\theta=\theta_1=\theta_2$, and particle 2 is a pure state $\mu_2=1$ and particle 1 is not, for different purities of particle 1 $\mu_1$ and phase rotations $\theta$. Plots for $\theta=0$ and $\mu=1$ and $\theta=\pi/4$ and $\mu=1$ are shown in Figs. \ref{1a} and \ref{1d} respectively. }
\label{fig:differentPurity}
\end{figure}

As a measure of entanglement we will employ the logarithmic negativity \cite{Vidal:2002}
\begin{align}
E_{\mathcal{N}} \left(\rho \right) := \log \left\| \rho^{\Gamma_A} \right\|_1,
\end{align}
where $\Gamma_A$ is the partial transpose and $\left\| \cdot \right\|_1$ denotes the trace norm, with  $\log(\cdot)$ denoting the natural logarithm.  The logarithmic negativity is a measure of the failure of the partial transpose of a quantum state to be a valid quantum state and is a faithful measure of entanglement for $1\times N$ mode Gaussian states \cite{Adesso:2004}.

For Gaussian states the logarithmic negativity is given by 
\begin{align}
E_{\mathcal{N}} := -\sum_k \log \tilde{v}_k \quad \forall \, \tilde{v}_k<1, \label{GaussianLogNegativity}
\end{align}
where $\left\{\tilde{v}_k\right\}$ is the symplectic spectrum of the partially transposed covariance matrix $\tilde{\mathbf{V}}$, i.e.  the eigenspectrum of $|i\boldsymbol{\Omega}\tilde{\mathbf{V}}|$.  The partial transpose of a covariance matrix is 
\begin{align}
\tilde{\mathbf{V}} = \boldsymbol{\theta}_{1|2} \mathbf{V} \boldsymbol{\theta}_{1|2},
\end{align}
where $\boldsymbol{\theta}_{1|2} = \diag (1,1,1,-1)$. 

We will use the logarithmic negativity to quantify the entanglement between the center of mass and relational degrees of freedom in $\mathbf{V}_{CMR} = \mathbf{M}_2 \mathbf{V}_{E} \mathbf{M}_2^T$, for $\mathbf{V}_E =  \mathbf{V}_1 \oplus \mathbf{V}_2 $, which corresponds to the two particles being in a factorized state $\rho_1\otimes\rho_2$ in the external partition. $\mathbf{V}_1$ and $\mathbf{V}_2$ will necessarily be of the form given in Eq. \eqref{singlemodecovaraince}. 

\begin{figure}
\subfloat[$\theta = 0$ and $\alpha =0$]{%
\includegraphics[width=0.25\textwidth]{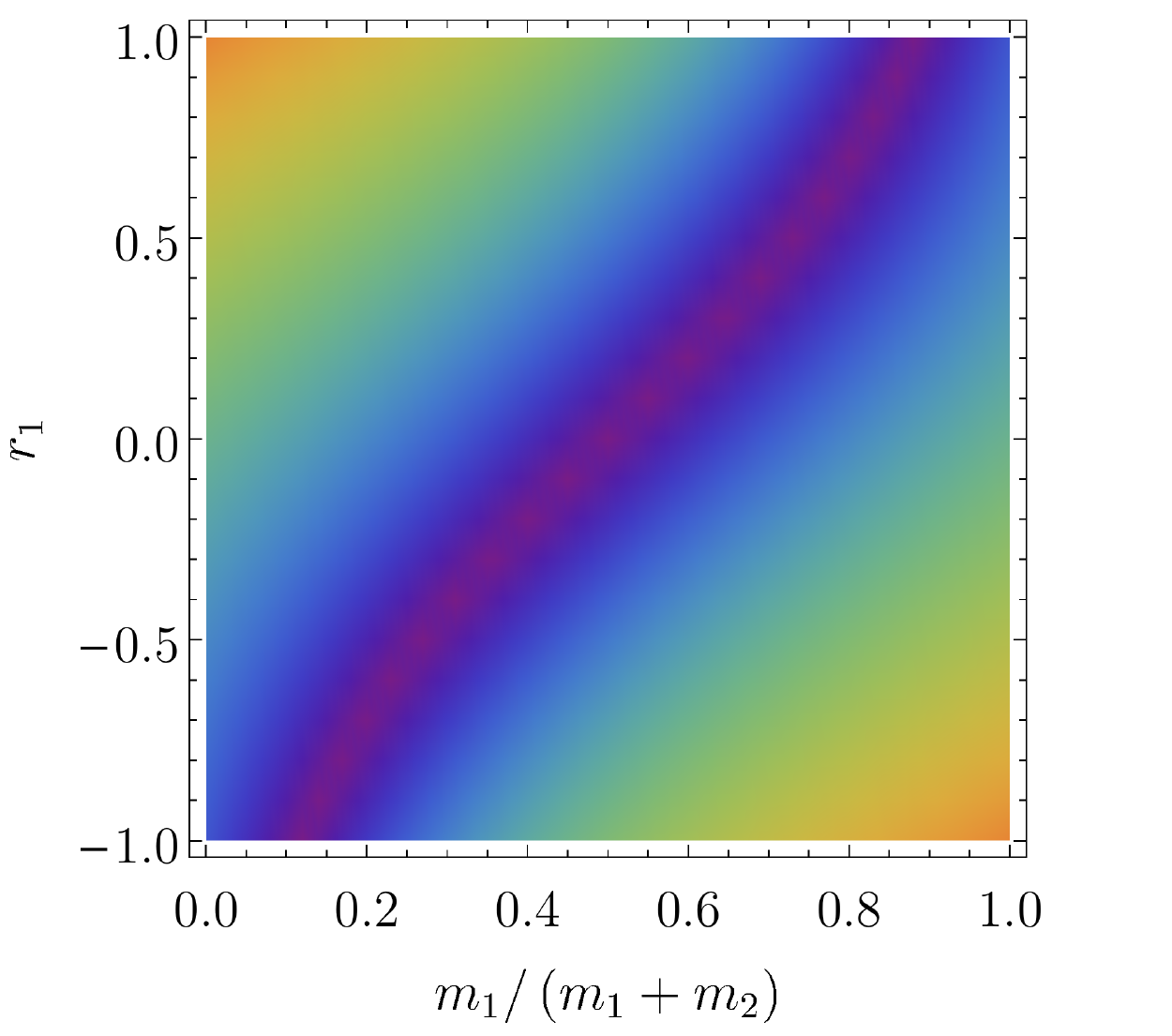}
}
\subfloat[$\theta = 0$ and $\alpha =0.5$]{%
\includegraphics[width=0.25\textwidth]{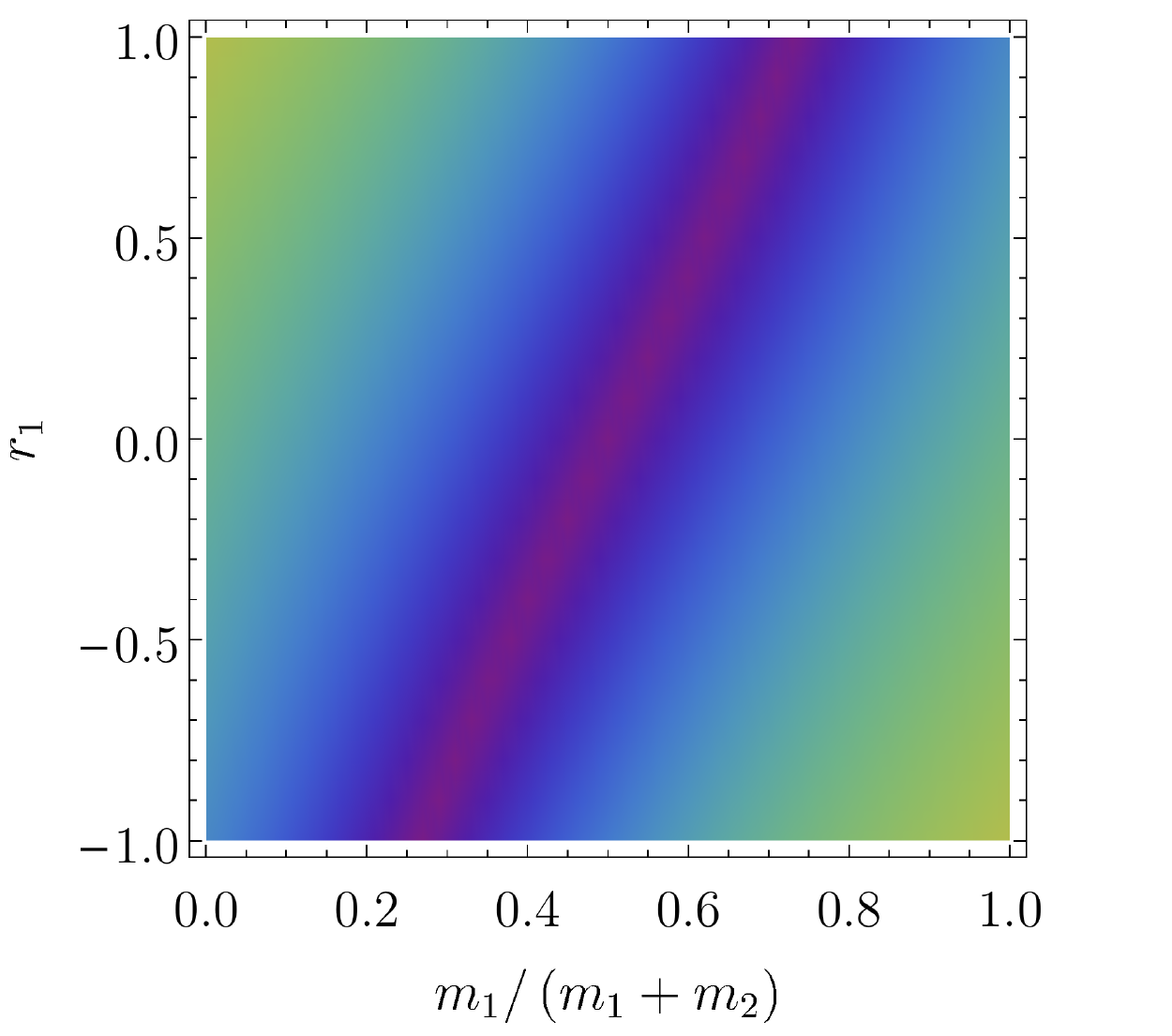}
}
\\
\subfloat[$\theta = \pi/4$ and $\alpha =0$]{%
\includegraphics[width=0.25\textwidth]{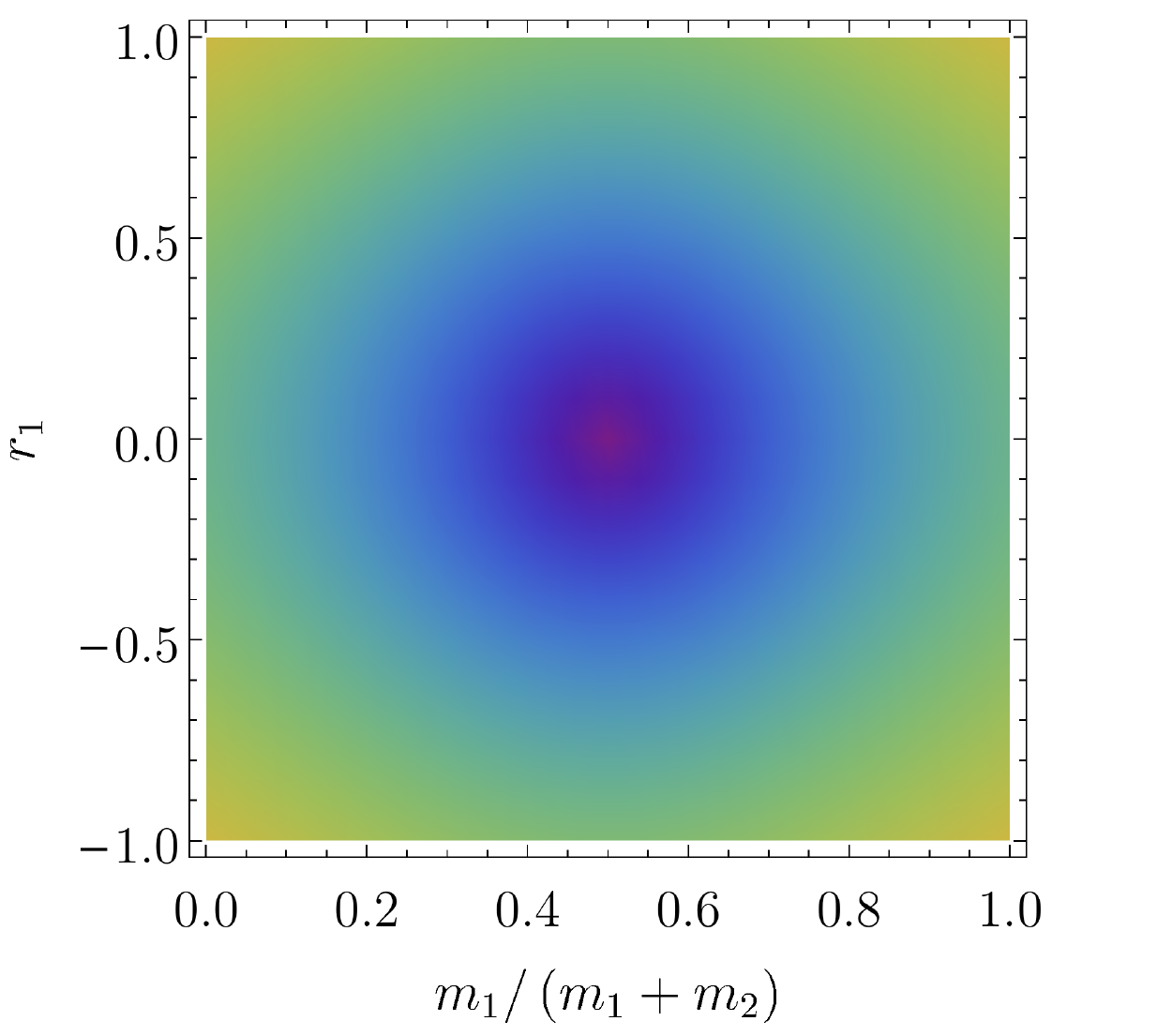}
}
\subfloat[$\theta = \pi/4$ and $\alpha =0.5$]{%
\includegraphics[width=0.25\textwidth]{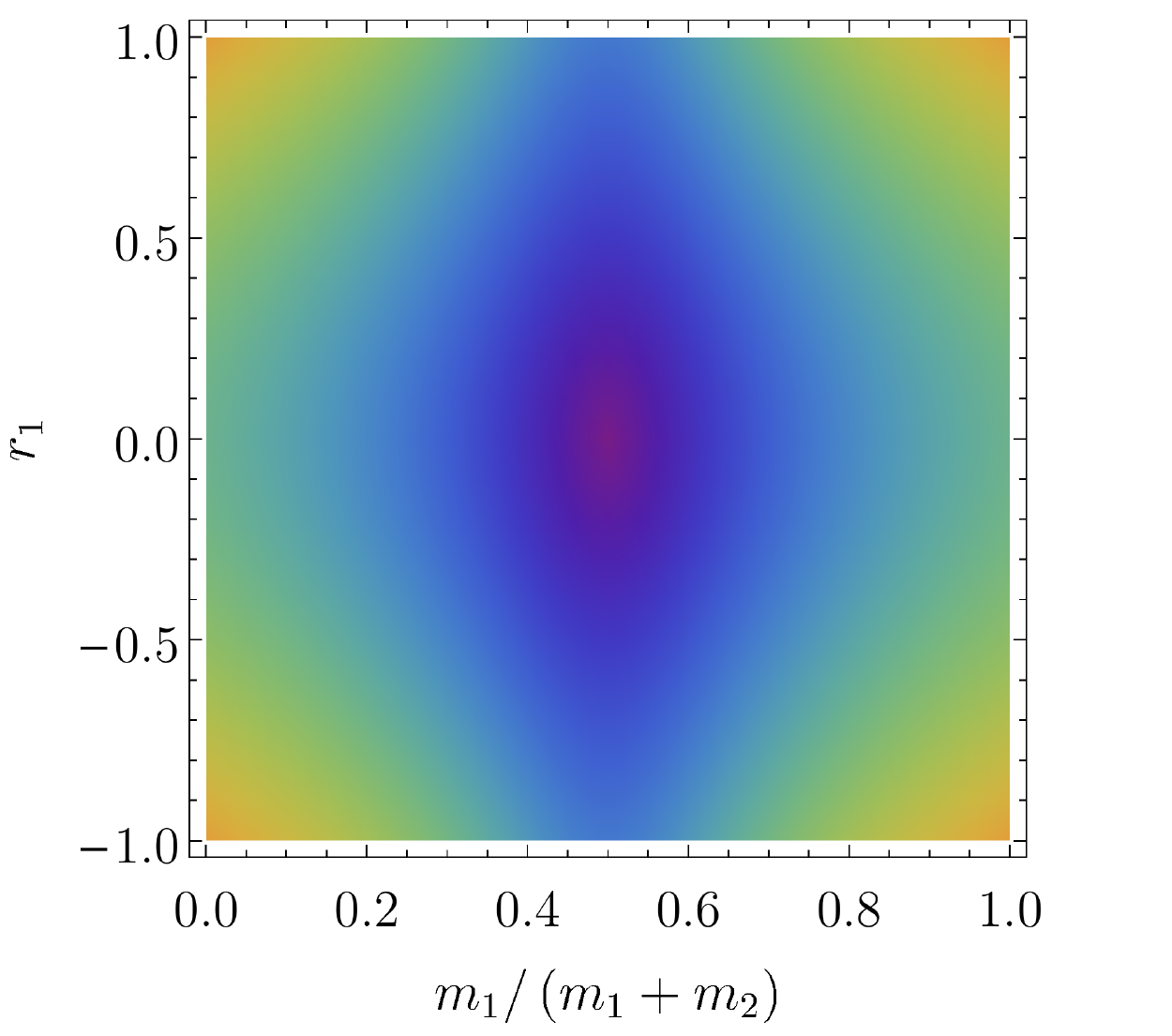}
}
\\
\subfloat{%
\includegraphics[width=0.25\textwidth]{legend.pdf}
}
\caption{(Colour online) The logarithmic negativity, as a measure of the entanglement between the center of mass and relation degrees of freedom, of the state associated with $\mathbf{V}_{CMR}$, when $\det \mathbf{V}_1 = \det \mathbf{V}_2 =1$ for $r_2 = \alpha r_1$, for different phase rotations $\theta = \theta_1 = \theta_2$ and values $\alpha$. Plots for $\theta=0$ and $\alpha=1$ and $\theta=\pi/4$ and $\alpha=1$ are shown in Figs.~\ref{1a} and \ref{1d} respectively.}
\label{fig:differentSqueezing}
\end{figure}

 Plots of the logarithmic negativity of the state associated with $\mathbf{V}_{CMR}$ for different choices of $\mathbf{V}_1$ and $\mathbf{V}_2$ are given in Figs.~\ref{fig:identicalV} (identical state parameters),  \ref{fig:differentPurity} (differing purity), and \ref{fig:differentSqueezing} (differing squeezing). Several trends emerge from a perusal of these figures.

We first note that equal-mass systems suppress entanglement between center of mass and relational degrees of freedom.  When particles in the external partition are prepared such that they have identical covariance matrices we find vanishing entanglement in the equal mass case regardless of the amount of squeezing and rotation. This occurs for both pure and mixed situations, respectively illustrated in Figs. \ref{fig:identicalV} and \ref{fig:differentPurity}.  As one of the masses gets larger, center of mass and relational entanglement increases for any fixed value of the squeezing parameter~$r$.  

The next trend we observe is that phase rotation, corresponding to squeezing along a rotated axis in phase space, appears to play a more important role than squeezing.  For a phase rotation $\theta=0$ we find that center of mass/relational entanglement is insensitive to the amount of squeezing.  As $\theta$ increases we see that squeezing plays an increasingly important role, particularly as the ratio of the masses increasingly departs from unity.  Not surprisingly, entanglement is greater for the pure case, shown in Fig.~\ref{fig:identicalV}, than for the mixed case, shown in Fig.~\ref{fig:differentPurity}.  

Asymmetric squeezing ($r_1 > r_2$), illustrated in Fig.~\ref{fig:differentSqueezing}, modifies this situation somewhat.  The zero-squeezing case in Figs.~ \ref{fig:differentSqueezing}a and \ref{fig:differentSqueezing}b,  shows vanishing entanglement when the masses are equal.  However there is increased center of mass/relational entanglement as the lighter particle is more strongly squeezed (Fig.~\ref{fig:differentSqueezing}a), a trend that is less pronounced as the differential squeezing decreases (Fig.~\ref{fig:differentSqueezing}b).  Vanishing  center of mass/relational entanglement takes place for increasingly larger values of the mass of the particle which is most squeezed.  Again we see that phase rotation plays a more significant role, restoring (in the maximal $\theta=\pi/4$ case) the symmetry present in the equal mass case (Figs.~\ref{fig:differentSqueezing}c and \ref{fig:differentSqueezing}d).  Here we see that a sufficient amount of differential squeezing can eliminate  center of mass/relational entanglement entirely (Fig.~ \ref{fig:differentSqueezing}c).

Decreasing the purity of the states of the particles in the external partition, shown in Fig.~\ref{fig:differentPurity}, indicates the same trends as for the pure case  (Fig.~\ref{fig:identicalV}). The main effects of decreased purity are to decrease the overall center of mass/relational entanglement and to widen the range of ratio of masses for which this entanglement vanishes.


\subsection{Three particles}
\label{Three particles} 

We consider now a similar analysis for a system of three particles with masses $m_1$, $m_2$, and $m_3$. When transforming a fully factorized state in the external partition $\mathcal{H} = \mathcal{H}_1\otimes\mathcal{H}_2\otimes\mathcal{H}_3$, to the center of mass and relational partition $\mathcal{H} = \mathcal{H}_{CM} \otimes\mathcal{H}_{R}$, there will again be entanglement generated between the center of mass and relational degrees of freedom. In addition, there will be entanglement generated among the relational degrees of freedom, a new feature not possible for the two particle system considered above.

The center of mass position and momentum operators, along with the relative position and momentum operators are again defined via Eq.~\eqref{CMRcoordinates}. The transformed covariance matrix is given by $\mathbf{V}_{CMR} =\mathbf{M}_3 \mathbf{V}_E \mathbf{M}_3^T$, where
\begin{align}
\mathbf{M}_3 :=
\begin{pmatrix}
\frac{m_1}{M} & 0 & \frac{m_2}{M}& 0 & \frac{m_3}{M}  & 0 \\
0 & 1 & 0 & 1 & 0 & 1 \\
-1 & 0 & 1 & 0 & 0 & 0 \\
0 & -\frac{m_2}{M} & 0& 1 -\frac{m_2}{M}& 0 &  -\frac{m_2}{M} \\
-1 & 0 & 0 & 0 & 1 & 0 \\
0 & -\frac{m_3}{M} & 0&  -\frac{m_3}{M}& 0 & 1  -\frac{m_3}{M}
\end{pmatrix}. \label{3particleM}
\end{align}

The relational state $\mathbf{V}_{23|1}$ of particles 2 and 3 as described by particle 1 is obtained by deleting the first and second rows and columns of $\mathbf{V}_{CMR}$. We observe that in the limit when $m_3$ vanishes and the columns and rows of $\mathbf{M}_3$ associated with particle 3 are deleted, that is the last two rows and columns, $\mathbf{M}_2$ as defined in Eq. \eqref{2particleM}  is recovered.

We assume the state of the three-particle system in the external partition is a fully factorized Gaussian state with the covariance matrix $\mathbf{V}_E =  \mathbf{V}_1 \oplus \mathbf{V}_2 \oplus \mathbf{V}_3$. For simplicity we restrict ourselves to the case when $\mathbf{V}_1=\mathbf{V}_2=\mathbf{V}_3$ and $\det \mathbf{V}_{E}=1$, in other words a pure state, with each of the three particles  identically squeezed in the same direction.

\begin{figure}
\subfloat[$\theta = 0$]{%
\includegraphics[width=0.25\textwidth]{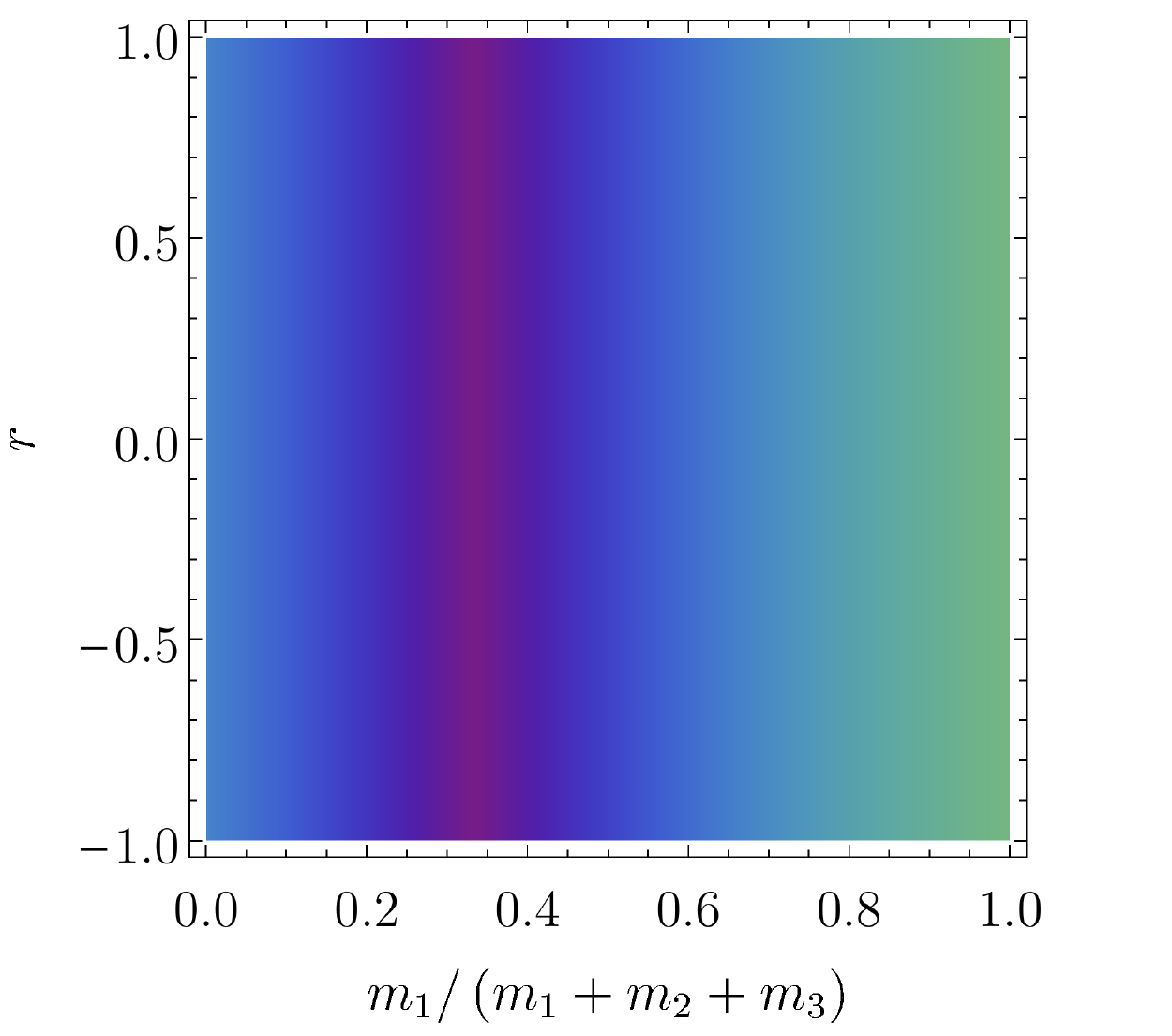}
}
\subfloat[$\theta = \pi/4$]{%
\includegraphics[width=0.25\textwidth]{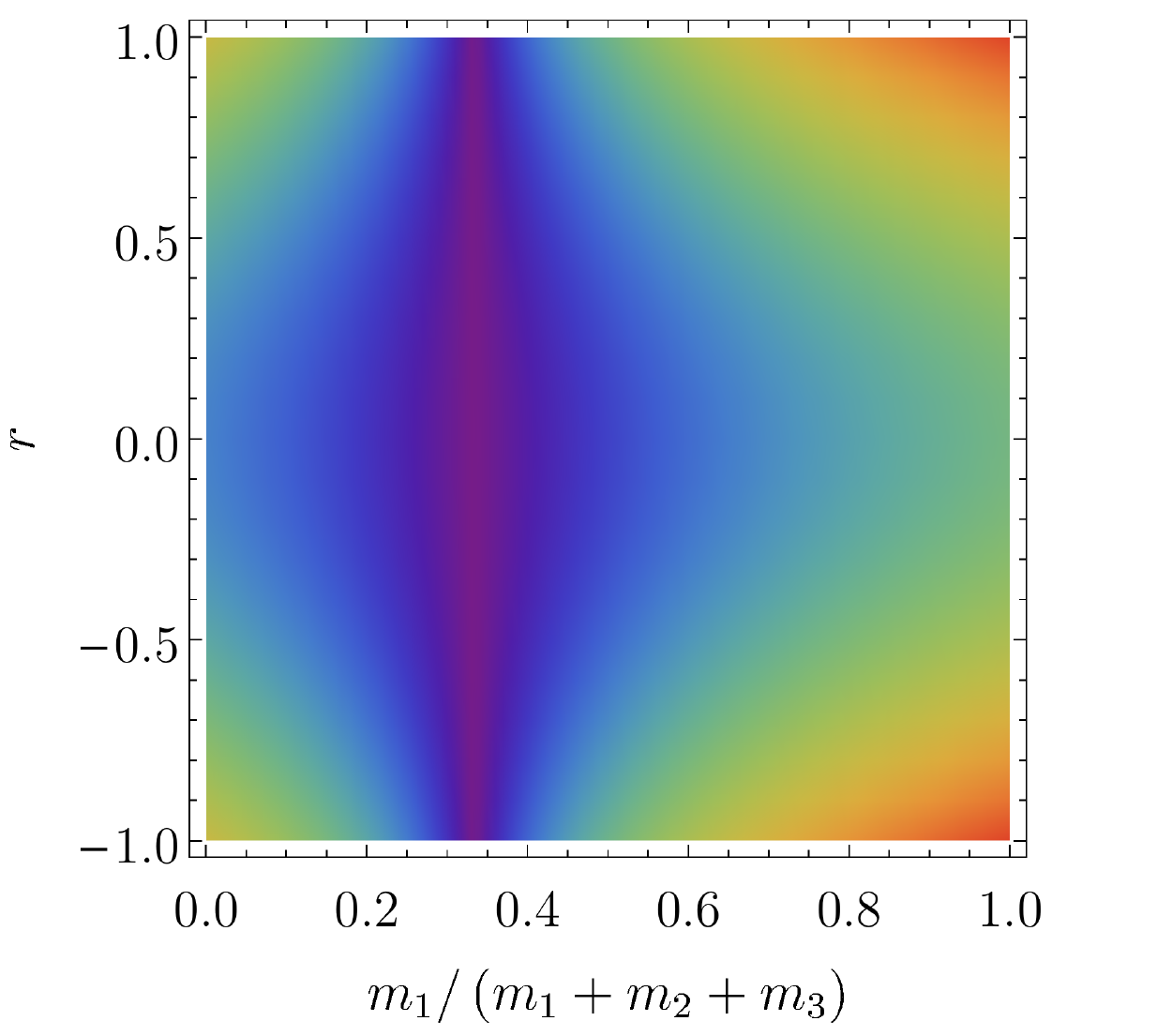}
}
\\
\subfloat[$\theta = 0$]{%
\includegraphics[width=0.25\textwidth]{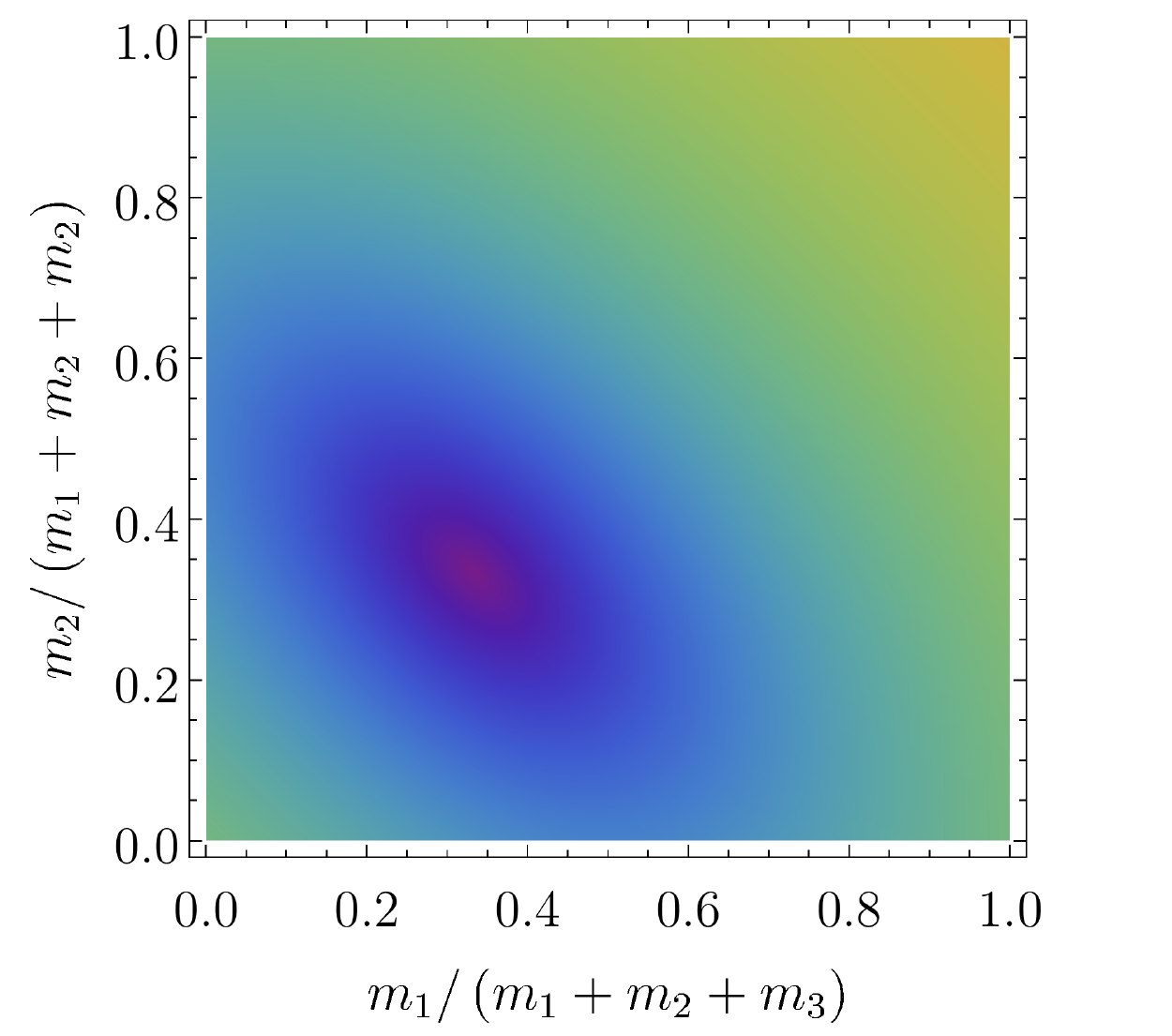}
}
\subfloat[$\theta = \pi/4$]{%
\includegraphics[width=0.25\textwidth]{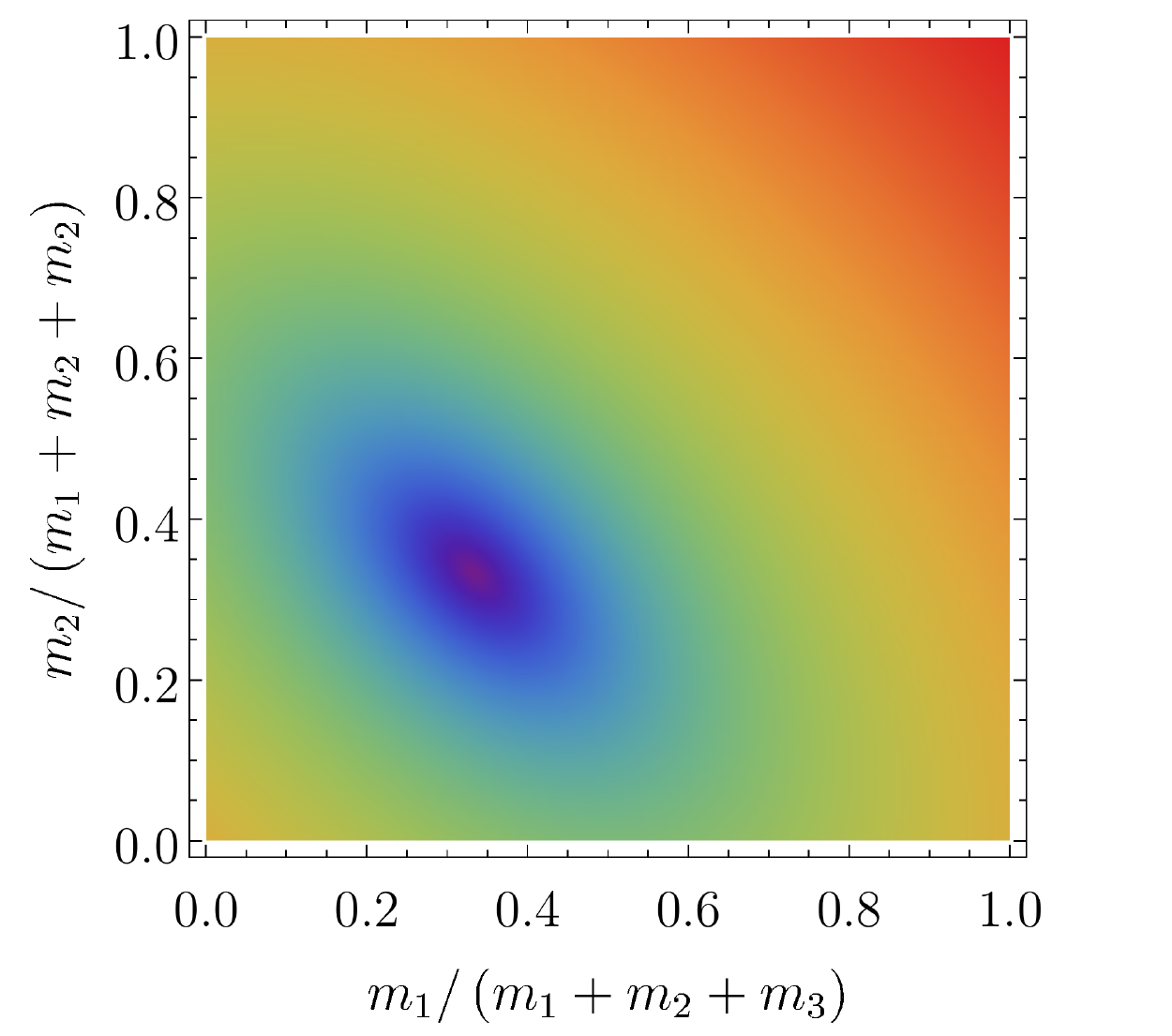}
}
\\
\subfloat{%
\includegraphics[width=0.25\textwidth]{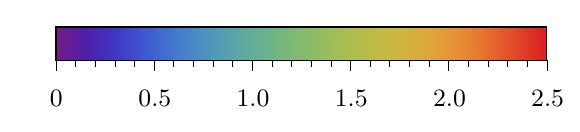}
}
\caption{(Colour online) The logarithmic negativity is plotted, as a measure of the entanglement between the center of mass and relation degrees of freedom, of the state associated with $\mathbf{V}_{CMR}$ describing three particles. In (a) and (b) the logarithmic negativity is plotted for the case when $m_2=m_3$, for various equal phase rotations $\theta_1=\theta_2=\theta_3=\theta$, as a function of the ratio $m_1/(m_1+m_2+m_3)$ and equal squeezing parameter $r_1=r_2=r_3=r$. In (c) and (d) the logarithmic negativity is  plotted  as a function of the two mass ratios $m_1/(m_1+m_2+m_3)$ and $m_2/(m_1+m_2+m_3)$ for various equal phase rotations $\theta$, with the equal squeezing parameter fixed at $r=0.7$. }
\label{fig:GlobelRelational3particle}
\end{figure}

\begin{figure}
\subfloat[$\theta = 0$]{%
\includegraphics[width=0.25\textwidth]{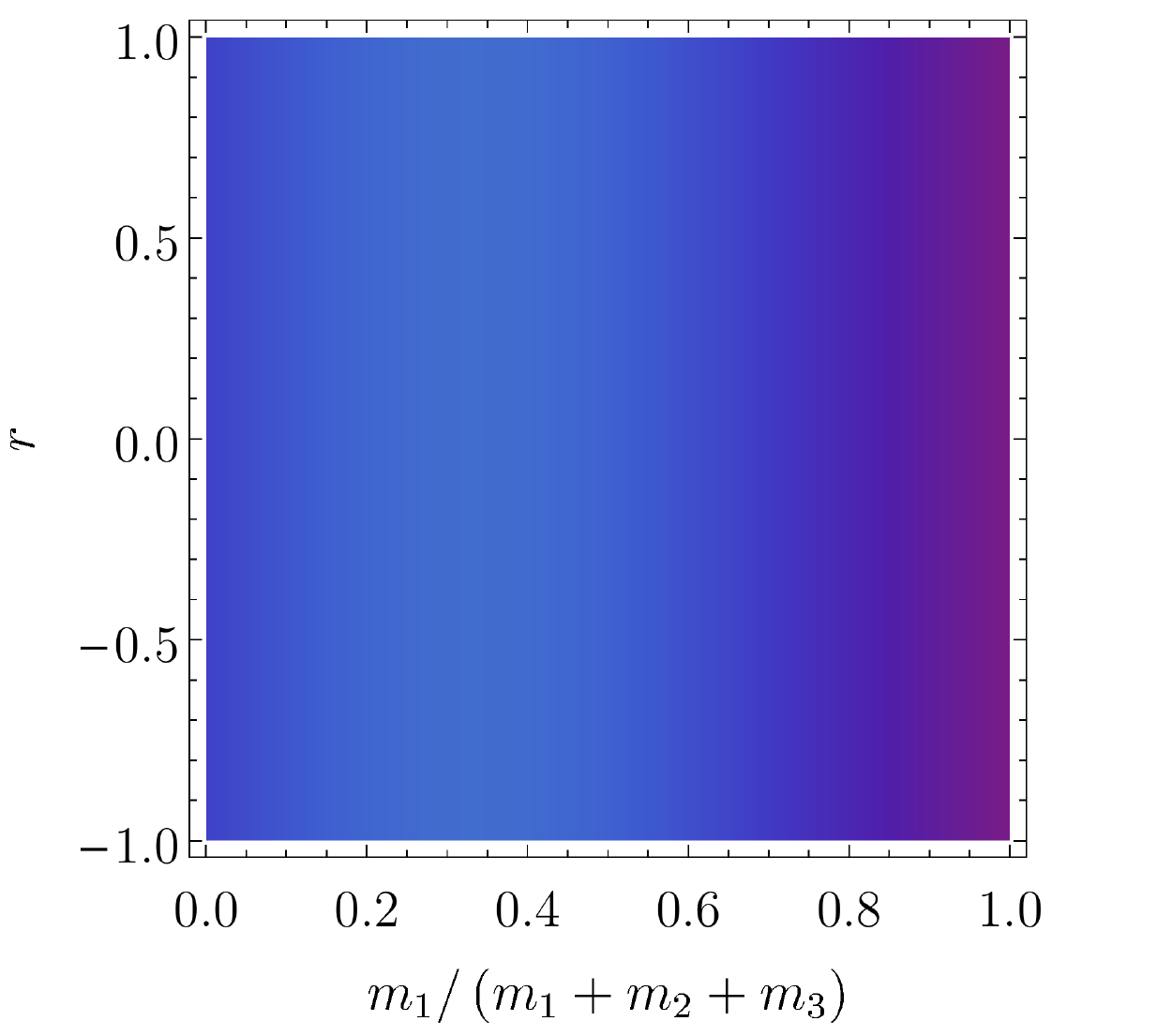}
}
\subfloat[$\theta = \pi/4$]{%
\includegraphics[width=0.25\textwidth]{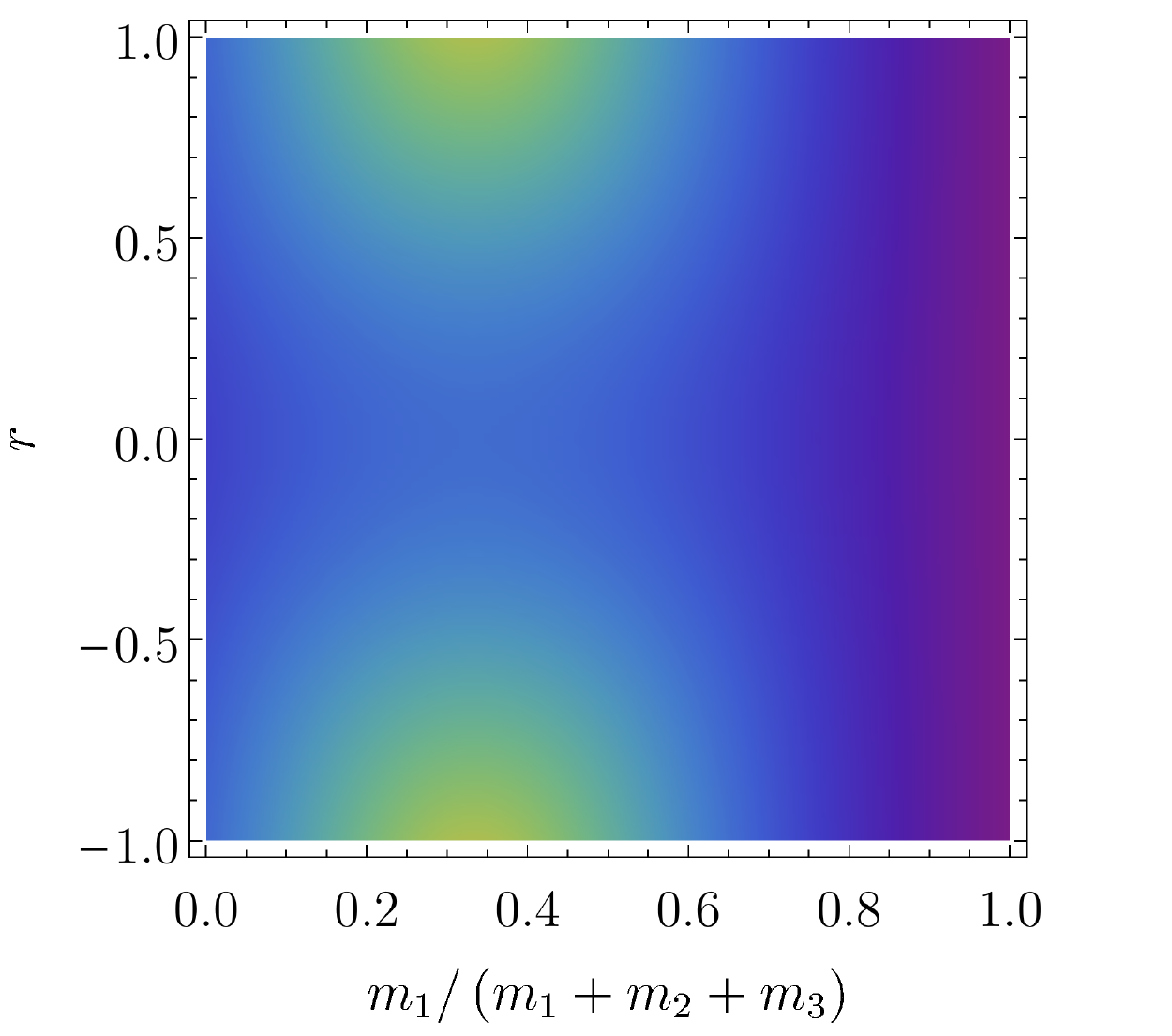}
}
\\
\subfloat[$\theta = 0$]{%
\includegraphics[width=0.25\textwidth]{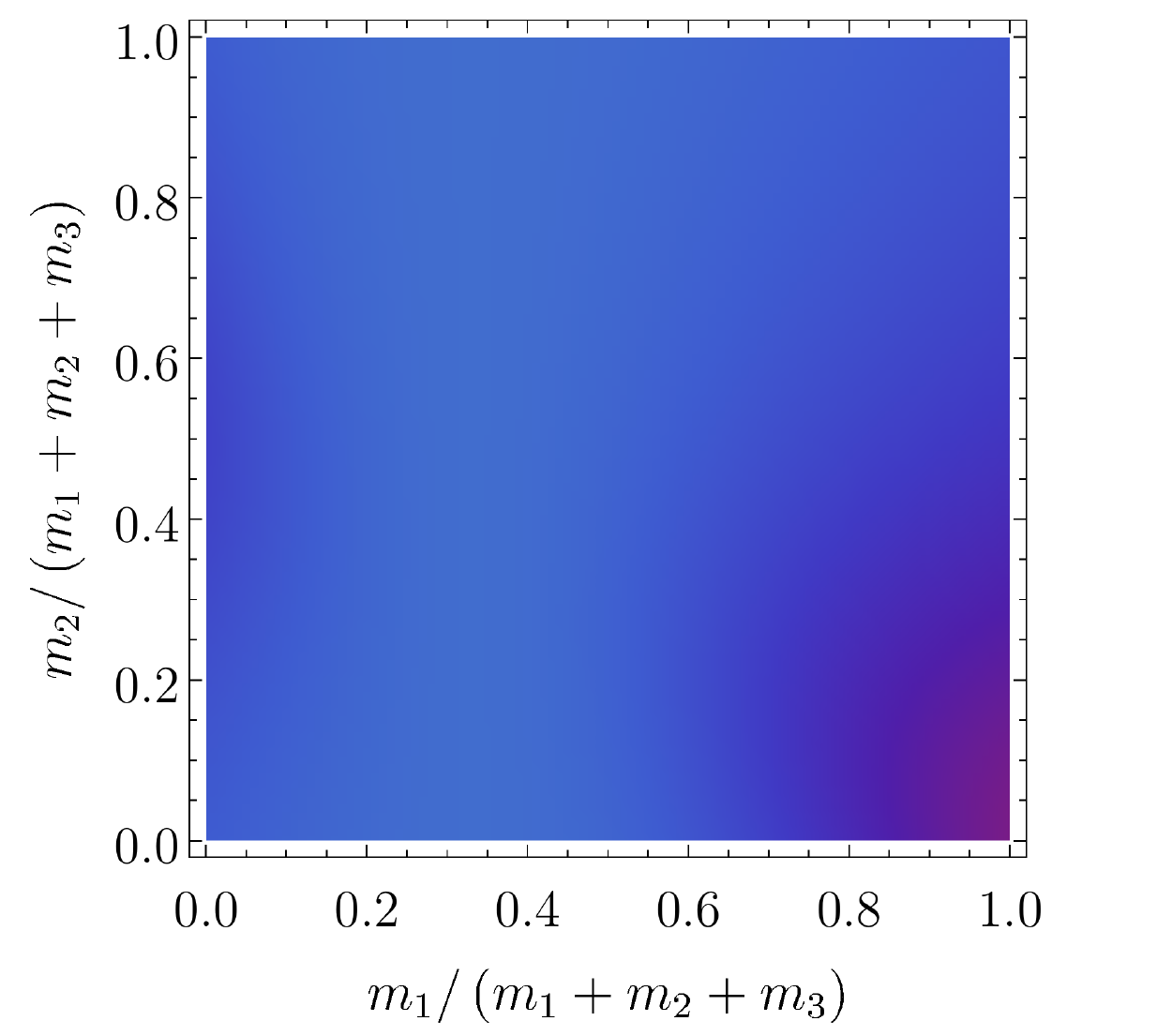}
}
\subfloat[$\theta = \pi/4$]{%
\includegraphics[width=0.25\textwidth]{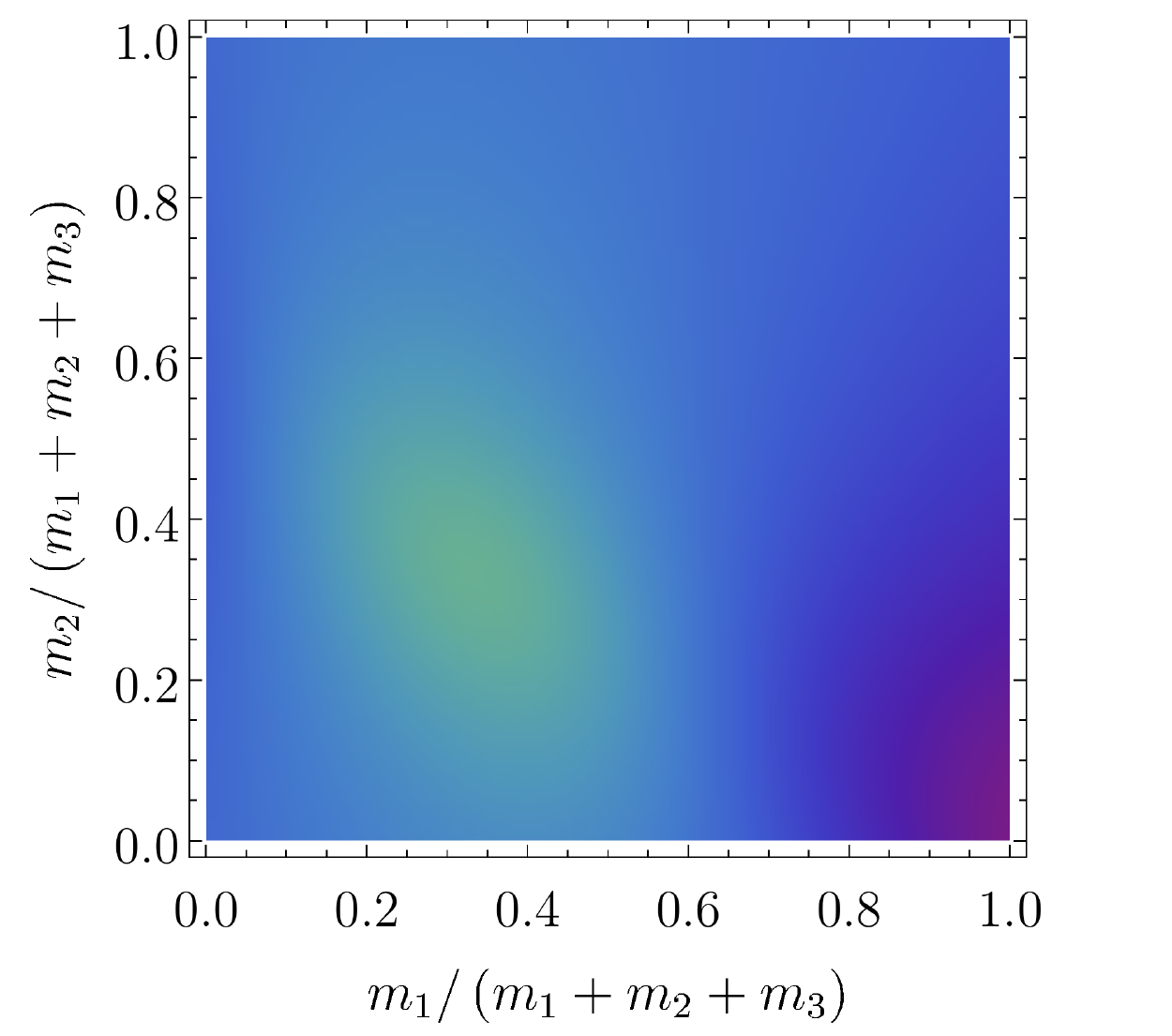}
}
\\
\subfloat{%
\includegraphics[width=0.25\textwidth]{legend2.pdf}
}
\caption{(Colour online) The logarithmic negativity of the relative state of particles 2 and 3 described by $\mathbf{V}_{23|1}$ is plotted, characterizing the entanglement among the relational degrees of freedom. In (a) and (b) the logarithmic negativity is plotted for the case $ m_2=m_3$, for various equal  phase rotations $\theta_1=\theta_2=\theta_3=\theta$, as a function of the ratio $m_1/(m_1+m_2+m_3)$ and equal squeezing parameters $r_1=r_2=r_3=r$. In (c) and (d) 
logarithmic negativity is  plotted as a function of the mass ratios $m_1/(m_1+m_2+m_3)$ and $m_2/(m_1+m_2+m_3)$ for equal squeezing parameter $r=0.7$ and for different equal phase rotations $\theta$.}
\label{fig:Relational3particle}
\end{figure}

In Fig.~\ref{fig:GlobelRelational3particle} the logarithmic negativity as a measure of entanglement between the center of mass and relational degrees of freedom in $\mathbf{V}_{CMR}$ is plotted for different choices of $\mathbf{V}_E$. In Fig.~\ref{fig:Relational3particle} the logarithmic negativity between the relational degrees of freedom in $\mathbf{V}_{23|1}$is plotted for different choices of $\mathbf{V}_E$.


We see similar trends for the center of mass/relational  entanglement as for the two-particle case, but qualitatively different behaviour of the internal-relational entanglement, i.e., the entanglement generated among the relational degrees of freedom---in the case at hand, the entanglement between particle 2 and 3 as described by particle 1.

The internal-relational entanglement, illustrated in Fig.~\ref{fig:Relational3particle}, shows strikingly different behaviour.  Such entanglement is maximized in the equal mass case, shown in Figs.~\ref{fig:Relational3particle}b and \ref{fig:Relational3particle}d provided there is some phase rotation.  In the absence of phase rotation, this effect vanishes.  For all values of the (equal) phase rotation parameter, we  observe that as the mass of the reference particle $m_1$ becomes infinite, the entanglement between particles 2 and 3 vanishes.  This is as expected, since this limit corresponds to particle 1 behaving as a classical reference frame with a large mass.
Indeed, we notice that in the limit $m_1\rightarrow\infty$, the $4\times4$ lower-right submatrix of $\mathbf{M}_3$ becomes the identity matrix, and the only effect of the change of coordinates is that of redefining the origin in space for the coordinates of the second and third particle.
 
\section{Discussion and outlook}
\label{Discussion}

We have highlighted issues involving quantum reference frames associated with non-compact groups. We began in Sec. \ref{Relational description for compact groups} by introducing the usually employed $G$-twirl as a relation description between quantum systems and demonstrated how it leads to unnormalized states when applied to non-compact groups. In Sec. \ref{Relational description for non-compact groups} we saw how the $G$-twirl over the group of translations and Galilean boosts leads to the appearance of the reduced state on the relational degrees of freedom previously considered by \citet{Angelo:2011}. We then examined the consequences of this relational description in Sec. \ref{Relational encoding and the translation group} by studying the entanglement that emerges between the center of mass degrees of freedom and the relational degrees of freedom, as well as the entanglement among the relational degrees of freedom, for a system of particles, when moving from a description of the quantum system entirely with respect to an external frame, to a description in which only the center of mass is specified with respect to an external frame and all other  degrees of freedom are relational. 

Two main observations emerged from studying the reduced state $\rho_R$ on the relational  degrees of freedom, introduced in Eq.~\eqref{RelationEncoding}, for systems of two and three particles. First, for fully separable Gaussian states in the external partition with identical second moments, entanglement between the center of mass degrees of freedom and relational degrees of freedom is minimized when the masses of the particles are the same. Second, again for fully separable Gaussian states in the external partition with identical second moments, in the limit when the mass of the reference particle, that is the particle for which the relational degrees of freedom are defined with respect to, becomes infinite, the entanglement among the relational degrees of freedom vanishes. This second observation suggests a meaningful way to interpret the external reference frame, with which we usually describe a quantum state with respect to, as the limit of a physical system, say a particle, in which its mass is taken to infinity \cite{Aharonov:1984}. The consequences of this second observation will be explored in future work.

The primary motivation for examining quantum reference frames associated with non-compact groups is to apply the quantum reference frame formalism to relativistic systems, in which the natural group associated with changes of a reference frame is the Poincar\'{e} group. It will be fruitful to explore to what extent the tools developed in this manuscript can be applied to the Poincar\'{e} group; however, one immediate obstacle is the problem of defining a covariant definition of the center of mass \cite{Aguilar:2013}.

Two other possible applications of the formalism introduced come to mind. The first is in constructing a relativity principle for quantum mechanics by studying changes of quantum reference frames, which was first suggested in Ref.~\cite{Palmer:2013}.  The second is to construct a relational quantum theory, similar to what was done in  Ref.~\cite{Poulin:2006}, for the Galilean group using the relational description in Eq. \eqref{RelationEncoding}, and examine how the usual ``non-relational'' theory emerges.

\begin{acknowledgments}
The authors would like to thank Nick Menicucci, Jacques Pienaar, and Daniel Terno for useful discussions. This work was  supported in part by the Natural Sciences and Engineering Research Council of Canada, by the European Union's Horizon 2020 research, and innovation programme under the Marie Sklodowska-Curie grant agreement No.~661338.  AS and MP acknowledge the hospitality of Macquarie University, where part of this work was conducted.
\end{acknowledgments}

\appendix*
\section{Purity of the relational state}
\label{Purity of the relational state}

The covariance matrices considered in sections \ref{Transforming to the global and relational partition} and \ref{Entanglement between the global and relational degrees of freedom} were of the form $\mathbf{V}_E = \mathbf{V}_1 \oplus \mathbf{V}_2$, where both $\mathbf{V}_1$ and $\mathbf{V}_2$ were given by Eq. \eqref{singlemodecovaraince}. The purity of $\mathbf{V}_{CMR}=\mathbf{M}_2 \mathbf{V}_E \mathbf{M}_2^T$ is given by
\begin{align}
\mu_{CMR} = \frac{1}{\sqrt{\det \mathbf{V}_{CMR}}} = \mu_1 \mu_2,
\end{align}
where $\mu_1$ and $\mu_2$ are the purities associated with $\mathbf{V}_1$ and $\mathbf{V}_2$ respectively.

The purity of the relational state $\mathbf{V}_{2|1}$ in Eq. \eqref{RelationalState21}, that is the state obtained from $\mathbf{V}_{CMR}$ by taking the partial trace over the center of mass degrees of freedom, is
\begin{align}
\mu_{2|1} =& \frac{1}{\sqrt{\det \mathbf{V}_{2|1}}} \nn \\
=& \mu_1 \mu_2 \Big[ \left.\mu_2^2 \tilde{m}_2^2 f_1^- f_1^+ +  \mu_1 \mu_2 \left(\tilde{m}_1^2 f_1^- f_2^+ + \tilde{m}_2^2 f_1^+ f_2^-\right) \right. \nn \\
&\left. \mu_1^2 \tilde{m}_1^2 f_2^- f_2^+   - \mu_2^2 \tilde{m}_2 ^2 g_1^2 + 2\mu_1 \mu_2 \tilde{m}_1 \tilde{m_2}g_1g_2  \right. \nn \\
& \left. - \mu_1^2 \tilde{m}_1^2 g_2^2  \right. \Big]^{-1/2},
\end{align}
where we have introduced the notation $\tilde{m}_i = m_i/(m_1+m_2)$.

If $\mathbf{V}_{CMR}$ is pure, which corresponds to both $\mathbf{V}_1$ and $\mathbf{V}_2$ being pure, then $\mu_{CMR}=1$ and $\mu_{2|1}$ is a genuine measure of entanglement between the center of mass and relational degrees of freedom. In this case, $\mu_{2|1}^{-2}$ simplifies to
\begin{align}
\mu_{2|1}^{-2} =&  \left( \tilde{m}_2 - \tilde{m_1} \right) \Big[ \sinh (2 r_1) \cosh (2 r_2) \cos (2 \theta_1) \nn \\
& - \sinh (2 r_2) \cosh (2 r_1) \cos (2 \theta_2) \Big]\nn \\
&+ (2 \tilde{m}_1 \tilde{m}_2+1) \cosh (2 r_1) \cosh (2 r_2) \nn \\
&-\sinh (2 r_1) \sinh (2 r_2) \Big[ 2 \tilde{m}_1 \tilde{m}_2 \cos (2 (\theta_1+\theta_2)) \nn \\
& +\cos (2 \theta_1) \cos (2 \theta_2) \Big]+  \tilde{m}_1^2 + \tilde{m}_2^2.
\end{align}

If the mass of the two particles are equal $m_1 = m_2$, $\mu_{2|1}^{-2}$ further simplifies to
\begin{align}
\mu_{2|1}^{-2} =& \frac{1}{4} \Big[-2 \sinh (2 r_1) \sinh (2 r_2) \cos (2 (\theta_1-\theta_2)) \nn \\
&+\cosh (2 (r_1-r_2))+\cosh (2 (r_1+r_2))+2 \Big].
\end{align}

For the case when $m_1 \neq m_2$, $r_1=r_2=r$ and $\theta_1 = \theta_2 = \theta$, corresponding to Fig.~\ref{fig:identicalV},  $\mu_{2|1}^{-2}$ becomes
\begin{align}
\mu_{2|1}^{-2} =&  2 \frac{m_1^2 + m_2^2}{\left(m_1 + m_2\right)^2} + \sin^2(2\theta)\nn \\
&\quad \times \left( \frac{m_1^2 + m_2^2}{\left(m_1 + m_2\right)^2}  \sinh^2 (2r) - 2 \frac{m_1m_2}{\left(m_1 + m_2\right)^2} \right) . \label{equalCase}
\end{align}
From Eq.~\eqref{equalCase},  we observe that when the masses of the two particles are identical $m_1=m_2$, the reduced state $\mathbf{V}_{2|1}$ is pure, i.e, $\mu_{2|1}=1$, which corresponds to vanishing entanglement between the center of mass and relational degrees of freedom in $\mathbf{V}_{CMR}$. This agrees with the plots of the logarithmic negativity in Fig.~\ref{fig:identicalV}. 

When the mass of either particle becomes infinite we find
\begin{align}
\mu_{2|1}^{-2} =&  2 + \sinh^2 (2 r) \cos^2 (2 \theta ).
\end{align}

\bibliography{QRF}
\end{document}